%% file: small_arxiv.tex
\documentclass[12pt]{article}
\usepackage{longtable}
\usepackage[section]{placeins}
\usepackage[top=1in, left=1in, right=1in, bottom=1in]{geometry}
\usepackage[parfill]{parskip}	
\usepackage{graphicx}
\usepackage{caption}
\usepackage{subcaption} 
\usepackage{mwe} 
\usepackage{cancel} 
\usepackage{soul} 
\usepackage[export]{adjustbox}
\usepackage{mathtools}
\usepackage{epstopdf}
\usepackage{bbm}
\usepackage{bm}
\usepackage{bbold}
\usepackage{upgreek}
\usepackage{amssymb}
\usepackage{amsmath}
\usepackage{xcolor}
\usepackage{amsfonts}

\usepackage{lscape}
\usepackage{enumitem}
\usepackage{rotating}
\usepackage{setspace}
\usepackage{diagbox}
\usepackage{listings}
\usepackage{threeparttable}
\usepackage{booktabs}
\usepackage[compact]{titlesec}
\usepackage{lmodern}
\fontfamily{lmtt}\selectfont
\usepackage[T1]{fontenc}

\usepackage[nocitation]{apacite}
\usepackage{natbib}
\bibpunct{(}{)}{;}{a}{}{,}
\usepackage{hyperref}
\usepackage{titling}
\usepackage{pifont}
\usepackage[capposition=top]{floatrow}
\newcommand{\subtitle}[1]{%
  \posttitle{%
    \par\end{center}
    \begin{center}\large#1\end{center}
    \vskip0.5em}%
}
\usepackage{amsthm}
\usepackage{multirow}
\usepackage{mathrsfs}
\usepackage{graphicx}
\usepackage{float}
\usepackage{stackengine}

\setstackgap{L}{0pt}
\usepackage[linesnumbered,ruled,vlined]{algorithm2e}
\SetKwInOut{Parameter}{parameter}
\SetKwInOut{Input}{input}

\newcommand{\rn}[1]{%
	\textup{\expandafter{\romannumeral#1}}%
}
\newcommand{\RN}[1]{%
	\textup{\uppercase\expandafter{\romannumeral#1}}%
}
\newcommand\independent{\protect\mathpalette{\protect\independenT}{\perp}}
\def\independenT#1#2{\mathrel{\rlap{$#1#2$}\mkern2mu{#1#2}}}

\makeatletter
\newcommand{\customlabel}[2]{%
   \protected@write \@auxout {}{\string \newlabel {#1}{{#2}{\thepage}{#2}{#1}{}} }%
   \hypertarget{#1}{#2}
}
\makeatother

\newcommand{\lt}{\ensuremath <}

\newcommand{\E}{\mathbb{E}}
\newcommand{\Pb}{\mathbb{P}}

\newcommand{\R}{\mathbb{R}}
\DeclareMathOperator*{\argmin}{arg\,min}
\DeclareMathOperator{\diag}{diag}

\DeclarePairedDelimiter\ceil{\lceil}{\rceil}

\newtheorem{remark}{Remark}

\newtheorem{theorem}{Theorem}[section]

\definecolor{dark-grey}{gray}{0.25}

\newcommand{\cmmnt}[1]{\ignorespaces}  

\begin{document}
\pagestyle{plain}





\makeatletter
\newcommand{\grande}{\bBigg@{2.25}}
\newcommand{\enorme}{\bBigg@{5}}

\newcommand{\blind}{0}

\newcommand{\tit}{Scalable kernel balancing weights in a nationwide observational study of hospital profit status and heart attack outcomes}

\if0\blind
{{\title{\tit\thanks{
This work was supported through a grant from the Alfred P. Sloan Foundation (G-2020-13946) and an award from the Patient Centered Outcomes Research Initiative (PCORI, ME-2022C1-
25648). The authors report there are no competing interests to declare. This work was completed while Kwangho Kim was a research associate at Harvard Medical School.
}}
\author{Kwangho Kim\thanks{Department of Statistics, Korea University, 145 Anam-ro, Seongbuk-gu, Seoul, 02841, South Korea; email: \url{kwanghk@korea.ac.kr}.} \and Bijan A. Niknam\thanks{Department of Biostatistics, Johns Hopkins Bloomberg School of Public Health, 615 N Wolfe St, Baltimore, MD 21205, United States; email: \url{bniknam1@jh.edu}.}\and Jos\'{e} R. Zubizarreta\thanks{Departments of Health Care Policy, Biostatistics, and Statistics, Harvard University, 180 Longwood Avenue, Office 307-D, Boston, MA 02115, United States; email: \url{zubizarreta@hcp.med.harvard.edu}.}
}}

\date{} 

\maketitle
}\fi

\if1\blind
\title{\tit}
\date{} 
\maketitle
\fi

\begin{abstract}
\input{small_sec0.tex}
\end{abstract}


\begin{center}
\end{center}
\clearpage
\doublespacing

\singlespacing
\pagebreak
\pagebreak
\doublespacing

\input{small_sec1.tex} 

\input{small_sec2.tex}

\input{small_sec3.tex}

\input{small_sec4.tex}

\input{small_sec5.tex}

\input{small_sec6.tex}

\input{small_sec7.tex}

\pagebreak

\bibliographystyle{asa}
\bibliography{bib2022}

\pagebreak
\setcounter{page}{1}
\singlespacing
\input{small_supp.tex}

\end{document}

%% file: small_sec0.tex
Weighting is a general and often-used method for statistical adjustment. Weighting has two objectives: first, to balance covariate distributions, and second, to ensure that the weights have minimal dispersion and thus produce a more stable estimator. A recent, increasingly common approach directly optimizes the weights toward these two objectives. However, this approach has not yet been feasible in large-scale datasets when investigators wish to flexibly balance general basis functions in an extended feature space. For example, many balancing approaches cannot scale to national-level health services research studies. To address this practical problem, we describe a scalable and flexible approach to weighting that integrates a basis expansion in a reproducing kernel Hilbert space with state-of-the-art convex optimization techniques. Specifically, we use the rank-restricted Nystr\"{o}m method to efficiently compute a kernel basis for balancing in {nearly} linear time and space, and then use the specialized first-order alternating direction method of multipliers to rapidly find the optimal weights. In an extensive simulation study, we provide new insights into the performance of weighting estimators in large datasets, showing that the proposed approach substantially outperforms others in terms of accuracy and speed. Finally, we use this weighting approach to conduct a national study of the relationship between hospital profit status and heart attack outcomes in a comprehensive dataset of 1.27 million patients. We find that for-profit hospitals use interventional cardiology to treat heart attacks at similar rates as other hospitals, but have higher mortality and readmission rates.

%% file: small_sec1.tex
\section{Introduction}


\subsection{Hospital profit status and heart attack care}

For-profit organizations, including hospitals, are widespread across the US healthcare system. The impact of profit motive on treatment and outcomes is a major focus of healthcare research. For-profit status may possibly incentivize hospitals to provide more efficient care, leading to better patient outcomes. For-profit hospitals may also invest more in technology and human capital to improve competitiveness and potentially patient outcomes. Alternatively, higher reimbursement for interventional treatment may possibly incentivize them to use invasive treatments more frequently than medically necessary. For example, Medicare pays hospitals more for heart attack patients who receive percutaneous coronary intervention (PCI), a treatment that is essential in some cases but of ambiguous benefit in others, potentially incentivizing unnecessary use. Previous studies of subgroups of hospitals reached conflicting conclusions on both PCI treatment rates and patient outcomes \citep{sloan2003does, shah2007impact}, emphasizing the need for a representative study of a comprehensive national sample.


\subsection{Weighting for covariate adjustment in large-scale datasets}

Covariate adjustment is essential in observational studies to remove bias due to differences in observed confounders. One method for adjustment is weighting, which uses each unit's probability of treatment assignment to form weighted samples that are free from differences due to observed confounders  \citep{austin2015moving}. Weighting attempts to remove bias by balancing observed covariate distributions while also producing robust and stable estimators. An appealing property of weighting is that it does not require explicit modeling of the outcome, and hence is part of the design stage of the study \citep{rubin2006design}. Further, a single set of weights can in principle be used to examine multiple outcomes \citep{little2019single}.

Procedurally, there are two broad approaches to weighting \citep{ben2021balancing}. 
The modeling approach emphasizes the fit of a model for the probability of treatment assignment given observed covariates, or propensity score \citep{rosenbaum1983}. However, covariate balance may be inadequate due to model misspecification or small samples; moreover, the weights may be highly variable and produce unstable estimators \citep{kang2007demystifying, zubizarreta2015stable}. 
To address these challenges, the balancing approach attempts to balance covariates directly, often by solving a convex optimization problem to find the weights of minimum dispersion that balance functions of the covariates (see, e.g., \citealt{hainmueller2012entropy}, \citealt{zubizarreta2015stable}\cmmnt{, \citealt{zhao2019covariate}; see also \citealt{imai2014covariate}, \citealt{athey2018approximate}, and \citealt{hirshberg2021augmented}}). This affords control over the form and degree of balance while also targeting estimator stability. Thus, the balancing approach can have better finite-sample performance \citep{chattopadhyay2020balancing}. 

Despite these appealing properties, the computational cost of many weighting algorithms prevents their scaling to the massive dimensions of many modern data sources in health sciences, which can reach millions of observations and hundreds of covariates. Moreover, to improve estimation accuracy, researchers increasingly recognize the importance of balancing richer covariate function classes beyond simple means and other low-dimensional parametric summaries. However, this additional complexity makes execution of weighting methods at scale even more challenging. 


\subsection{Contribution and outline}

We propose a weighting approach that addresses practical barriers to the use of balancing methods in big data. Following \cite{wong2018kernel} and \cite{hazlett2020kernel}, we adopt a kernel balancing approach which assumes the outcome regression functions are in a flexible function space represented by a kernel. Further, we use the {recently developed} rank-restricted Nystr\"{o}m approximation \citep{wang2019scalable} that gives a provable approximate solution to the efficient kernel basis calculation problem. This allows us to construct the kernel basis functions in {near-}linear time and space, with strong relative-error guarantees. We then incorporate these functions into a quadratic program {with} linear constraints, which are efficiently solved via a specialized first-order alternating direction method of multipliers \citep{osqp2020}, yielding the stable kernel balancing weights of minimum dispersion that approximately balance {the covariate functions} \citep{zubizarreta2015stable}. We analyze the worst-case bias bound of the resulting linear estimator and discuss its implications for parameter {selection}. We benchmark the proposed weighting estimator's performance against common alternatives in extensive simulation studies, providing new insights into how weighting estimators perform at scale, finding that the proposed approach considerably outperforms others in terms of both accuracy and speed. Finally, we use the new approach in a national study of heart attack treatment and outcomes by hospital profit status.

The paper outline is as follows. Section 2 describes the setup, the balancing approach to weighting, and kernel balancing. Section 3 details the rank-restricted Nystr\"{o}m  approximation, the operator splitting solver for quadratic programs, and the proposed scalable kernel balancing approach integrating the two concepts, along with an analysis of the worst-case bias bound for the resulting weighting estimator. Section 4 outlines the simulation study settings, examines the performance of the proposed estimator, and provides general insights into the speed and accuracy of various weighting estimators. Section 5 examines the relationship between hospital profit status and heart attack treatment and outcomes in a national Medicare dataset. Section 6 concludes.

%% file: small_sec2.tex
\section{Estimation framework} 

\subsection{Two approaches to weighting}\label{subsec:two_approaches}

Consider an observational study with $n$ triplets $(X, A, Y) \overset{iid}{\sim} \Pb$, where $X \in \mathcal{X} \subset \mathbb{R}^d$ is a vector of observed covariates, $A$ is a binary treatment assignment indicator with the number of treated units $n_t=\sum_{i=1}^n A_i$ and control units $n_c = n - n_t$, and $Y \in \mathbb{R}$ is a real-valued outcome. 
Under the potential outcomes framework for causal inference \citep{rubin1974estimating}, we write $Y_i = A_i Y_i^{1} + (1-A_i)Y_i^{0}$, where $Y_i^{a}$ is the potential outcome for unit $i$ under treatment $A = a$; $a = 0, 1$. In our case study, $A=1$ if a patient was admitted to a for-profit hospital, and $A=0$ otherwise. 
Here, we wish to estimate the \textit{average treatment effect on the treated} (ATT)\footnote{While we focus on estimating the ATT, our discussion also applies to the \textit{average treatment effect} (ATE), and more generally to the \textit{target average treatment effect} (TATE), $\E_{\mathbb{T}}(Y^1 - Y^0)$, where $\mathbb{T}$ represents the measure of the target population of interest \citep{kern2016assessing}. 
The ATE and ATT are special cases of the TATE where $\mathbb{T} = \Pb$ and $\mathbb{T} = \Pb(A = 1)$, respectively.},
\noindent
\begin{equation} \label{eqn:def-att}
\E(Y^1 - Y^0 \mid A = 1). 
\end{equation}
\noindent
In order to identify {the ATT} from observational data, we invoke the following assumptions \citep{rosenbaum1983}:
\noindent

{\scshape Assumption 2.1:} (i) \textit{Consistency}, $Y = Y^a$ if $A = a$; (ii) \textit{Exchangeability}, $A \independent Y^0 \mid X$; \linebreak
(iii) \textit{Positivity}, $\Pb(A = 0 \mid X = x) > 0$ a.s. $[\Pb]$.

Given these assumptions, the ATT in \eqref{eqn:def-att} is identified by
\begin{align*} 
\E(Y \mid A = 1) - \E\{\E(Y \mid X, A = 0) \mid A = 1\} \equiv \E(Y \mid A = 1) - \psi. 
\end{align*}
The first term \cmmnt{represents the mean potential outcomes of the treated sample under treatment and }can be estimated easily by taking a sample average. 
However, estimating $\psi$, the mean conditional potential outcomes of the treated under control, requires not only the aforementioned assumptions but also careful covariate adjustment. 

A simple yet general weighting estimator for $\psi$ is given by
\begin{equation} \label{eqn:linear-estimator}
\widehat{\psi} = \sum_{\{i\mid A=0\}}\hat{w}_i Y_i \quad \text{for} \quad \hat{w} \geq 0.    
\end{equation}
where $\hat{w}=(\hat{w}_1,\ldots,\hat{w}_{n_c}) \in \R^{n_c}$ is the set of estimated weights that removes imbalances in covariate distributions such that after weighting adjustment, the control sample resembles the treated sample in aggregate.

In the {modeling approach} to weighting, we explicitly model the unknown propensity score $\pi(X)=\Pb(A=1 \mid X)$, and then construct the weights $\hat{w}$ using the estimated $\widehat{\pi}$. \cmmnt{We define a normalized version of the classic {Horvitz-Thompson} estimator as}One of the most widely-used modeling-based weighting estimators is the \emph{Hajek estimator} defined by
\begin{align} \label{eqn:modeling-Hajek}
\widehat{\psi}_{\text{mod}} = \frac{1}{n}\sum_{i=1}^n\left\{\frac{\widehat{\pi}(X_i)(1-A_i)}{1-\widehat{\pi}(X_i)}Y_i\right\}\Bigg/\frac{1}{n}\sum_{i=1}^n\left\{\frac{\widehat{\pi}(X_i)(1-A_i)}{1-\widehat{\pi}(X_i)}\right\}.    
\end{align}
The performance of model-based weighting estimators hinges upon the accuracy of estimation of $\widehat{\pi}$. Often, flexible nonparametric models are employed for $\widehat{\pi}$, but these approaches encounter difficulties when scaled to very large datasets \citep[e.g.,][]{zhou2017machine}.
\cmmnt{As noted earlier, the performance of model-based weighting estimators hinges upon the accuracy of estimation of $\widehat{\pi}$. If the true $\pi$ is correctly estimated using a parametric model (e.g., logistic regression), asymptotic normality is attained with fast $\sqrt{n}$ rates; however, strong functional form assumptions must be satisfied. In contrast, a more flexible nonparametric model requires weaker assumptions on $\pi$. However, slower convergence rates alongside  computationally expensive algorithms and hyperparameter optimization processes \citep{zhou2017machine} pose challenges for scalability of nonparametric methods to very large datasets.}

\cmmnt{The balancing approach to weighting is a more recent alternative approach which avoids explicit modeling of $\pi$ and rather directly optimizes for the weights $w_i$ that satisfy certain criteria specified by the investigator.  This approach can directly target not only covariate balance, but also the dispersion of the weights, thereby yielding a more stable estimator. Consequently, the balancing approach has been found to provide better finite-sample performance than the modeling approach \citep{chattopadhyay2020balancing}.}
The balancing approach to weighting is a more recent alternative approach which avoids explicit modeling of $\pi$ and rather directly optimizes for the weights. This approach can directly target not only covariate balance, but also the dispersion of the weights, thereby yielding a more stable estimator. Consequently, the balancing approach has been found to provide better finite-sample performance than the modeling approach \citep{chattopadhyay2020balancing}. Here, we focus on the stable balancing weights (SBW) \citep{zubizarreta2015stable}, which solve the following quadratic program
\cmmnt{Popular balancing approaches include \citet{hainmueller2012entropy}, \citet{zubizarreta2015stable}, and \citet{chan2016globally}. In particular, \citet{zubizarreta2015stable} developed the stable balancing weights (SBW), which solve the following quadratic program}
\begin{equation}
\label{eqn:stable-balancing-weights}
\begin{aligned}
    & \underset{w \in \mathbb{R}^{n_c}}{\text{minimize}} \quad \sum_{\{i\mid A=0\}}(w_i-\bar{w})^2\\
    & \text{subject to } \quad \left\vert \sum_{\{i\mid A=0\}}w_iB_b(X_i) - \frac{1}{n_t}\sum_{\{i\mid A=1\}}B_b(X_i)\right\vert \leq \delta, \quad b= 1, 2, ..., M, \\
    & \phantom{\text{subject to }} \quad \, \sum_{\{i\mid A=0\}}w_i = 1, \quad w \geq 0,
\end{aligned}    
\end{equation}
\noindent where $\bar{w} = \frac{1}{n_c} \sum_{\{i\mid A=0\}}w_i$,  $B_1,...,B_M$ are $M$ real-valued basis functions of the covariates that span the model space in which the unknown outcome regression function {is assumed to} lie, and $\delta$ is a prespecified tolerance level for the maximum acceptable imbalance. We use $\widehat{\psi}_{\text{bal}}$ to denote the weighting estimator when the set of weights $\hat{w}$ represents the solution of \eqref{eqn:stable-balancing-weights}. We focus on the stable balancing weights for two reasons. First, the objective function targets the variance of the weights and hence \eqref{eqn:stable-balancing-weights} is a convex quadratic program (QP) to which we may apply a state-of-the-art algorithm to improve scalability. Second, the optimal weights have minimum variance and balance the covariates, thereby controlling the bias and variance of the resulting estimator, and achieving the semiparametric efficiency bound under certain regularity conditions \citep{wang2020minimal}.
Further, approximate rather than exact balancing makes the approach's performance less sensitive to settings with limited covariate overlap. The parameter $\delta$ controls the trade-offs between covariate balance and the variance of the weights, and therefore the bias-variance trade-off of the resulting estimator. A smaller value of $\delta$ reduces imbalances and therefore bias due to differences in observed covariates, while a larger $\delta$ produces a set of weights with smaller variance and therefore a more stable estimator. In practice, subject-matter experts are typically consulted to ensure the choice of $\delta$ is suitable for the application at hand. However, some recent work has explored how to choose $\delta$ in an empirical fashion. For example, appealing to asymptotic theory, \citet{hirshberg2020augmented} and \citet{hirshberg2021minimax} recommend selecting $\delta$ in the context of the dual formulation of the weighting problem and setting it to be roughly equivalent to the conditional variance of the outcome. Other automatic selection approaches have also been suggested \citep{kallus2020, wang2020minimal, zhao2019covariate}; however, more theoretical and empirical work remains to fully characterize their performance.

\cmmnt{
Although appearing distinct, the balancing and modeling approaches to weighting are in fact connected. For example, \citet{tan2020regularized, wang2020minimal} showed that the balancing procedure is equivalent to the penalized regression problem with a certain link function. \citet{zhao2016entropy} also showed that entropy balancing is doubly robust.
More recently, \citet{chattopadhyay2021implied} established a connection between the balancing weights procedure and the implied weights of outcome linear regression models.
}

\subsection{Kernel balancing}
\cmmnt{Many plausible outcome models relating baseline covariates to heart attack outcomes exist, yet the true model is unknown.}
Let $\mu_0(X)=\E(Y^0 \mid X)$ be the outcome regression function in some model space $\mathcal{M}$. It can be shown that the bias of the weighting estimator \eqref{eqn:linear-estimator} with the weights $\hat{w}$ such that
\begin{align} \label{eqn:outcome-model-balancing}
    \max_{\mu_0 \in \mathcal{M}} \left\vert \sum_{\{i\mid A=0\}}\hat{w}_i\mu_0(X_i) - \frac{1}{n_t}\sum_{\{i\mid A=1\}}\mu_0(X_i)\right\vert \leq \delta
\end{align}
is on the order of $\delta$ \citep{ben2021balancing}.
Hence, to maximize the chance that the weighting adjustment eliminates the covariate imbalance in $\mu_0$, it is desirable to choose a flexible set of basis functions $\{B_1, ..., B_M\}$ that spans a general model space for $\mu_0$.
One such space is the Reproducing Kernel Hilbert Space (RKHS). For a Mercer kernel $K: \mathcal{X} \times \mathcal{X} \rightarrow \R$ and a given sample of size $n$, the RKHS $\mathcal{H}_K$ is defined by the completion of the function space 
\[
    \left\{ f: f(x) = \sum_{i=1}^n \alpha_i K(X_i,x), \, \alpha_i \in \R \right\}
\]
with respect to the norm $\Vert f \Vert_{\mathcal{H}_K}=\sqrt{\sum_{i=1}^n\sum_{j=1}^n\alpha_i\alpha_j K(X_i,X_j)}=\sqrt{\alpha^\top \bm{K} \alpha}$, $\forall f \in \mathcal{H}_K$, where $\alpha = (\alpha_1, \ldots,$ $\alpha_n)^\top \in \R^n$ and $\bm{K}$ is the $n \times n$ kernel matrix in our observational data with $\bm{K}_{ij}= K(X_i,X_j)$. For a clearer conceptualization of the function space $\mathcal{H}_K $, define a feature map $\Phi(x): x \mapsto [\sqrt{\lambda_1}\varphi_1(x), \sqrt{\lambda_2}\varphi_2(x), \ldots]$, where $\{\lambda_j\}$ and $ \{\varphi_j\}$ are, respectively, the eigenvalues and orthonormal eigenfunctions of the kernel operator such that $\int K(x,y)\varphi_j(y)dy=\lambda_j\varphi_j(x)$. Then by Mercer’s Theorem it can be shown that $\Phi(x)^\top \Phi(y) = K(x,y)$, and that for any $f \in \mathcal{H}_K$, $f$ can be expanded either in terms of $K$ or in terms of the bases $\{\varphi_j(\cdot)\}_{j=1}^\infty$, such that 
\begin{align} \label{eqn:spectral-representation}
    f(x)=\sum_{i=1}^n \alpha_i K(X_i,x)=\sum_{j=1}^\infty \beta_j \varphi_j(x)
\end{align}
for some $\alpha_i,\beta_j \in \R$. In the kernel balancing approach, we assume that $\mu_0 \in \mathcal{H}_K$ as follows.

{\scshape Assumption 2.2}:
$\mathcal{M} = \mathcal{H}_K$ for a Mercer kernel $K$ and its induced RKHS $\mathcal{H}_K$.

Letting the balancing constraints in \eqref{eqn:stable-balancing-weights} be $B_b(\cdot)=\varphi_b(\cdot)$, $b=1,\ldots, \infty$, we pursue approximate mean balance on the basis set for $\mathcal{H}_K$, $\{\varphi_j(\cdot)\}_{j=1}^\infty$, to control bias according to \eqref{eqn:outcome-model-balancing}. 
By \eqref{eqn:spectral-representation}, this is equivalent to using $B_b(\cdot)=\bm{K}_{b}^\top$, $b=1,\ldots, n$, where $\bm{K}_{i}$ denotes the $i$-th row of $\bm{K}$. 
In other words, one may use the $n$ columns of $\bm{K}$ as bases for $\mu_0$ rather than $\{\varphi_j(\cdot)\}_{j=1}^\infty$. Note that $\bm{K}$ can be constructed without computing $\varphi_b$.
Achieving mean balance on columns of $\bm{K}$ can also guarantee covariate balance in  $\widehat{\mu}_0$. If we find $\widehat{\mu}_0$ as the solution to the regularized problem
\begin{align} \label{eqn:regularization-model}
    \underset{\mu_0 \in \mathcal{H}_K}{\min} \sum_{i=1}^n \mathcal{L}(\mu_0(X_i), Y^0_i) + \tau \Omega(\Vert \mu_0 \Vert_{\mathcal{H}_K}),
\end{align}
for some loss function $\mathcal{L}:\R^2 \rightarrow \R$, monotone increasing function $\Omega$, and $\tau > 0${,} then by the representer theorem \citep{wahba1990spline}, $\widehat{\mu}_0$ always has the form
$
    \widehat{\mu}_0(x) = \sum_{i=1}^n \alpha^*_i K(X_i, x) = \sum_{j=1}^\infty {\beta}^*_j \varphi_j(x) 
$
for some $\alpha^*_i, {\beta}^*_j \in \R$. Hence, the same argument applies as above.

However, incorporating $n$ balancing constraints into the optimization is challenging for large $n$. 
Admitting the spectral decomposition, one may write $\bm{K} = U \Lambda U^\top$ where $\Lambda$ is the diagonal matrix with a set of eigenvalues in decreasing order. Then under Assumption (2.2), the bias of the linear estimator \eqref{eqn:linear-estimator} due to imbalance on $\bm{K}$ is given by
\begin{align} \label{eqn:att-bias}
    \text{Bias}(\widehat{\psi}; \bm{K}) \coloneqq 
    C_0\left(\widehat{w}^\top U_c - \frac{1}{n_t} \mathbb{1}_{n_t}^\top U_t\right) \Lambda U^\top \alpha_0,
\end{align}
for some $C_0 > 0$ and $\alpha_0$ such that $\Vert \mu_0 \Vert_{\mathcal{H}_K}=\sqrt{\alpha_0^\top \bm{K} \alpha_0}$, where $U_c$ ($U_t$) are rows of $U$ corresponding to the control (treated) units and $\mathbb{1}_{n_t}$ is the vector of ones of length $n_t$. Based on this finding, \citet{hazlett2020kernel} proposed balancing the first $r \ll n$ eigenvectors of $\bm{K}$, i.e., the first $r$ columns of $U$, rather than balancing all $n$ columns of $\bm{K}$. While easier than balancing all $n$ columns, this can still be prohibitive due to the high computational cost of the eigenvalue decomposition; standard algorithms for the eigenvalue decomposition of a dense $n \times n$ matrix require $O(n^2)$-space and have $O(n^3)$-time complexity. In the next section, we discuss an effective approximation method for $\bm{K}$ and a scalable strategy for the corresponding basis function construction.

\begin{remark}[Positivity under a model class] 
    Although positivity is not strictly required for identification under the assumption that $\mathcal{M} = \mathcal{H}_K$, it may play a role in estimation by influencing the feasibility of {finding} the weights in the balancing problem. At the extreme, one may not feasibly find weights if no control units have similar rows of $\bm{K}$ as a given treated unit. However, the notion of positivity in this setting is more subtle than the standard positivity assumption, as there are several factors simultaneously affecting the feasibility of weighting solutions, including the type of kernel, bandwidth, tolerance level, and the observed sample at hand.
\end{remark}

\begin{remark}[Considerations for Kernel selection] 
    For a given a kernel $K$, the corresponding RKHS is unique.
    One desirable property for $K$ is \emph{universal approximation}, which allows us to approximate an arbitrary continuous function uniformly on the feature space induced by $\Phi$ \citep{micchelli2006universal}. Following \citet{hazlett2020kernel}, we use the Gaussian kernel, which satisfies the universal approximation property. However, other kernels could also be considered depending on prior knowledge of the properties of the outcome regression function (see \citeauthor[2006, Section 3,]{micchelli2006universal} for examples of universal kernels). When setting the kernel bandwidth, we likewise follow \citet{hazlett2020kernel} by setting it to the column rank of $X$. Like the type of kernel, the chosen bandwidth may also affect both the performance and feasibility of the kernel balancing estimator. The optimal choices of the kernel and its bandwidth are important considerations in causal inference that warrant further investigation.
\end{remark}

%% file: small_sec3.tex
\section{Toward scalable and flexible weighting}

\subsection{Scalable kernel basis calculation using the rank-restricted Nystr\"{o}m approximation} \label{subsec:kernel-basis}

The kernel approach provides a flexible way to balance functions of the covariates. However, its expensive computational cost presently prevents its use in large datasets.
This problem can be surmounted with low-rank matrix approximations. Rather than compute the eigenvalue decomposition of the entire kernel, one may instead perform a partial singular value decomposition (SVD) \citep{lehoucq1998arpack}. However, the performance of this approach depends heavily on the singular spectrum structure of $\bm{K}$, and only realizes computational gains when $\bm{K}$ is sparse. Another alternative is to use a class of randomized algorithms such as the randomized QR decomposition or randomized SVD \citep{halko2011finding}, but despite having good convergence, these algorithms still generally require $O(n^2)$ time.   

A more general alternative is the Nystr\"{o}m method \citep[e.g.,][]{gittens2016revisiting}, an efficient technique for low-rank approximation of a full kernel matrix based on a subset of its columns. The method can substantially reduce the time complexity since only a portion of $\bm{K}$ is sampled for computation. However, when $n$ is extremely large, even a small matrix may no longer be small in practical terms, and the accuracy of the matrix approximation drastically declines if the number of sampled columns is too small. In this context, \citet{pourkamali2018randomized} note the limitations of traditional Nystr\"{o}m methods for very large data sets, a problem we address using a more recently introduced form of the Nystr\"{o}m method. 

We adopt the rank-restricted Nystr\"{o}m method recently proposed by \citet{wang2019scalable}. Unlike traditional Nystr\"{o}m methods, the rank-restricted method allows the target rank to be much smaller than the number of sampled columns while still providing guarantees on the relative-error bound. For $\bm{K} \in \R^{n \times n}$, we first sample a set of $m \ll n$ column indices $\mathcal{I}_m$ (with the magnitude of sampled columns $m$ also termed a \emph{sketch size}) . While various sampling or \emph{sketching} schemes have been proposed \citep[see, e.g.,][and Table \ref{tbl:sketch-size-sampling}]{kumar2012sampling}, we use uniform sampling here as its minimal computational cost per sampled column guarantees $O(n)$ time and hence has advantages for scalability. After sampling columns of $\bm{K}$, we form a matrix $C \in \R^{n \times c}$ with $C_{ij}=K(X_i, X_j)$, $i \in \{1,...,n\}$, $j \in \mathcal{I}_m$, and a matrix $W \in \R^{m \times m}$ with $W_{ij}=K(X_i, X_j)$, $i,j \in \mathcal{I}_m$. Then, the standard Nystr\"{o}m approximation for $\bm{K}$ is
\begin{align} \label{eqn:standard-Nystrom}
    \widetilde{\bm{K}} = C W^+ C^\top,
\end{align}
\noindent where $W^+$ is the Moore-Penrose (pseudo) inverse of $W$. 

\begin{table}[!t]
\centering
\begin{tabular}{ccc}
\hline
Sketching method    & Sketch size ($c$) &  Time complexity\\ \hline 
Uniform sampling    & $O\left(\frac{s\mu}{\epsilon} + s\mu\log s \right)$ & $O(nc)$\\ 
Leverage sampling   & $O\left(\frac{s}{\epsilon} + s\log s \right)$ & $O(n^2s)$ \\ 
Gaussian projection & $O\left(\frac{s}{\epsilon}\right)$ & $O(n^2c)$\\ 
Subsampled randomized Hadamard transform                & $O\left( \left(s + \log n\right)\left(\epsilon^{-1} + \log s \right) \right)$ & $O(n^2\log c)$\\ 
CountSketch         &  $O\left(\frac{s}{\epsilon} + s^2 \right)$ & $O(n^2)$ \\ \hline
\end{tabular}
\caption{Sufficient sketch sizes for popular sketching methods to construct the proposed Nystr\"{o}m approximation $\left(\widetilde{\bm{K}}\right)_s$ \citep[][Lemma 10]{wang2019scalable}. Here, $\epsilon \in (0,1)$ and $\mu \in \left[1, \frac{n}{s}\right]$ denote the error parameter and the row coherence of $V_s$, respectively, where ${V}_s{\Sigma}_s{V}_s^\top$ is the truncated SVD of $\bm{K}_s$.}
\label{tbl:sketch-size-sampling}
\end{table}

Because small singular values in $W$ often result in numerical instability and thus large errors in $W^+$, we mitigate this problem by adopting a commonly used regularization heuristic. Specifically, for a \emph{regularization parameter} $l < m$, we compute the SVD of $W$ as $U_W\Lambda_W U_W^\top$ and let $W_l=U_{W,l} \Lambda_{W,l} U_{W,l}^\top$ be the truncated SVD of $W$, where $\diag(\sigma_1, ..., \sigma_l) = \Lambda_{W,l} \in \R^{l \times l}$. Then, we use
\begin{align} \label{eqn:Nystrom-approx-K}
    \widetilde{\bm{K}}_l = CW_l^{-1}C = C U_{W,l} \Lambda_{W,l}^{-1} U_{W,l}^\top C^\top
\end{align}
as a regularized version of $\widetilde{\bm{K}}$.  

Finally, we compute the rank-restricted Nystr\"{o}m approximation with a \emph{target rank} $s$ as
\begin{align} \label{eqn:DR-Nystrom-approx-K}
    (\widetilde{\bm{K}}_l)_s = DD^\top \quad \text{where} \quad D = C U_{W,l} \Lambda_{W,l}^{-1/2} \tilde{V}_{R,s} \equiv R\tilde{V}_{R,s} \in \R^{n\times s},
\end{align}
where $\tilde{V}_{R,s}$ is the dominant $s$ right singular vectors of $R = C U_{W,l} \Lambda_{W,l}^{-1/2}$, $s < l$.

Because the rank-restricted Nystr\"{o}m approximation offers an effective way to construct the $s$ most relevant balancing constraints, the approach computes the weights faster while also yielding a more stable estimator.
With uniform sampling, the time and space complexities to compute $(\widetilde{\bm{K}}_l)_s$ are $O(m^3 + nml)$ and $O(nm + m^2)$, respectively. {Since $m$, a proxy for our computational resources, increases at much slower rates than $n$ in typical settings,} both are now {near-}linear in $n$. Hence, \eqref{eqn:DR-Nystrom-approx-K} approximates the eigenvector-based kernel bases with major computing gains. This method is also much more computationally efficient in large sample sizes than the rank-$s$ randomized SVD, whose time complexity is $O(n^2s + s^3)$ \citep{li2014large}.



\subsection{Scalable balancing weights using OSQP} \label{subsec:fast-optimization}

Having described a practical way of computing the kernel representation, we next address optimization of the weights at scale. For large-scale convex optimization, the \emph{alternating direction method of multipliers} (ADMM) is likely the most common solution method due to its computational efficiency \citep{boyd2011distributed}. However, ADMM has some limitations, such as its inability to detect infeasibility, the sensitivity of its performance to parameter selection and data setting, and its diminished speed advantage on high-accuracy settings. Recently, \cite{osqp2020} developed the operator splitting solver for quadratic programs (OSQP), a state-of-the-art ADMM-based solver for general QPs that addresses these problems.

The OSQP algorithm is particularly well-suited to solving for the stable balancing weights in \eqref{eqn:stable-balancing-weights}. First, the most time-intensive step of OSQP is solving the linear system of equations (Line 9 in Algorithm \ref{algo:kernel-osqp}); however, since the coefficient matrix of the linear system is symmetric quasi-definite and sparse (with sparsity $\approx \frac{1}{n_c}$), a number of efficient algorithms can be applied to this step, such as the QDLDL factorization \citep{davis2005algorithm}. Second, the coefficient matrix of the linear system is always non-singular regardless of the step-size parameter values, preserving numerical stability without sacrificing speed. Finally, when careful tuning of $\delta$ is desired (for example, by using the bootstrap \citep{chattopadhyay2020balancing, wang2020minimal}), OSQP can execute this step much more efficiently through factorization caching and warm starting. See Appendix B for more details on ADMM and OSQP.
\cmmnt{Specifically, although the tuning procedure requires a number of iterations, because the coefficient matrix in the relevant linear system is fixed across iterations and does not depend on $\delta$, factorization caching can be used, storing the factorization from the first iteration for use in subsequent ones. Also, because small changes in $\delta$ do not substantially affect weighting solutions, warm starting can be used for added efficiency. See Appendix B for more details on ADMM and OSQP.}


\subsection{Implementation} \label{subsec:implementation}

This section describes an algorithm to formulate the proposed estimator incorporating kernel-based stable balancing weights.
Assume we have obtained the rank-$s$ Nystr\"{o}m approximation $(\widetilde{\bm{K}}_l)_s=DD^\top$ as in \eqref{eqn:DR-Nystrom-approx-K}, and for the matrix $D$, let $D_t \in \R^{n_t \times s}$ and $D_c \in \R^{n_c \times s}$ be the rows of $D$ corresponding to the treated and control units, respectively. Given these inputs, the proposed kernel-based stable balancing weights are given by the optimal solution to the following quadratic program:
\begin{equation} \label{eqn:kernel-SBW-QP}
\begin{aligned}
    & \underset{\substack{w \in \R^{n_c} \\ z \in R^{1+n_c+l}}}{\text{minimize}} \quad w^\top  w  - \frac{1}{n_c}w^\top \mathbbm{1}_{n_c}\\
    & \text{subject to} \quad Qw = z, z \in \mathcal{C}(\delta),
\end{aligned}    
\end{equation}
for $
Q = 
\begin{bmatrix}
\mathbbm{1}_{n_c} & \mathbb{I}_{n_c} & D_c
\end{bmatrix}^\top
$ and a set $\mathcal{C}(\delta) = \left\{z \mid l(\delta) \leq z \leq u(\delta) \right\}$ where
\begin{align*}
    l(\delta)=\begin{bmatrix}1 & \mathbb{0}_{n_c} & \overline{D_t} - \delta
\end{bmatrix}^\top, \quad u(\delta) = \begin{bmatrix}1 & \mathbbm{1}_{n_c} & \overline{D_t}  + \delta \end{bmatrix}^\top, \quad \delta=\left(\delta_1, ... ,\delta_s \right).
\end{align*}

Here, $\mathbb{I}_r \in \R^{r\times r}$ is the $r\times r$ identity matrix, and $\mathbbm{1}_r, \mathbb{0}_r \in \R^{1 \times r}$, the vectors of ones and zeros of length $r$, respectively; $\delta$ is a vector of maximum allowable imbalances specified by the investigator, and $\overline{D_t}\in \R^{1 \times s}$ is a row vector containing the mean of each column of $D_t$. 
Program \eqref{eqn:kernel-SBW-QP} can be efficiently solved via OSQP. Algorithm \ref{algo:kernel-osqp} describes computation of the proposed estimator in detail. Note that the kernel balance constraints can also be used alongside traditional balance requirements for low-dimensional summaries of the covariates, such as their means. 

Program \eqref{eqn:kernel-SBW-QP} expresses the stable balancing weights problem \eqref{eqn:stable-balancing-weights} in the standard nomenclature of ADMM, with the approximated kernel basis $D_{ib}$ as $B_b(X_i)$. 
Computation of $D$ is informed by the scaling parameters $m,l,s$. The available computational capacity should be considered when selecting the sketch size $m$. Choices for the regularization parameter $l$ and the target rank $s$ are related to the relative-error bound analysis in the next section. 

\begin{algorithm}[!t]
\caption{Fast stable kernel balancing weights}
\label{algo:kernel-osqp}
\textbf{input:} Sample $Z_1,...,Z_n$, vector $\delta$, integers $s <l < m \ll n$, and kernel $K(\cdot, \cdot)$\\
Sample a set $\mathcal{I}_m$ of $m \ll n$ indices in $\{1,...,n\}$ \\
Form $C \in \R^{n \times m}$ with $C_{ij}=K(X_i, X_j)$, $i \in \{1,...,n\}$, $j \in \mathcal{I}_m$, and $W \in \R^{m \times m}$ with $W_{ij}=K(X_i, X_j)$, $i,j \in \mathcal{I}_m$\\
Compute the rank-$l$ truncated SVD of $W$ as $W_l=U_{W,l} \Lambda_{W,l} U_{W,l}^\top$ \\
Compute the rank-$s$ truncated SVD of $R = C U_{W,l} \Lambda_{W,l}^{-1/2}$ as $\tilde{U}_{R,s} \tilde{\Lambda}_{R,s} \tilde{V}_{R,s}^\top$, then compute $D = R \tilde{V}_{R,s}$ \\
Compute $Q, l(\delta), u(\delta)$ using \eqref{eqn:kernel-SBW-QP} \\
Choose parameters $\rho > 0$, $\sigma > 0$, and $\upalpha \in (0,2)$;
initialize $w^0$, $y^0$, $z^0$, and $\nu^0$; and set $k=0$\\
\Repeat{termination criterion satisfied}{
$(\widetilde{w}^{k+1}, {\nu}^{k+1}) \leftarrow$ solve 
$\begin{bmatrix}
-(1+\sigma)\mathbb{I}_{n_c} & -Q^\top  \\
-Q & \rho^{-1}\mathbb{I}_{1+n_c+s}
\end{bmatrix}
\begin{bmatrix}
\widetilde{w}^{k+1}   \\
\nu^{k+1} 
\end{bmatrix}
=
-\begin{bmatrix}
\sigma w^{k} - 1/{n_c}\mathbbm{1}_{n_c}   \\
z^{k} - \rho^{-1}y^k
\end{bmatrix}$ \\
$\widetilde{z}^{k+1} \leftarrow z^k + \rho^{-1}(\nu^{k+1} - y^k)$ \\
$w^{k+1} \leftarrow \upalpha \widetilde{w}^{k+1} + (1-\upalpha)w^k$ \\
$z^{k+1} \leftarrow \Pi_{\mathcal{C}(\delta)}\left(\upalpha\widetilde{z}^{k+1} + (1-\upalpha)z^k + \rho^{-1}y^k \right)$ \\
$y^{k+1} \leftarrow y^k + \rho\left(\upalpha \widetilde{z}^{k+1} + (1-\upalpha)z^k - z^{k+1} \right)$\\
$w \leftarrow w_k$, $k \leftarrow k+1$
}
\textbf{output:} $\sum_{i=1}^{n_c} w_i Y_{c,i}$, where $Y_c$ is an outcome vector for the control units
\end{algorithm}

\subsection{Analysis of the worst-case bias bound}

In the next theorem, we analyze a worst-case bound on the bias,  $\text{Bias}(\widehat{\psi}; \bm{K})$, in \eqref{eqn:att-bias}: 
\begin{align*} 
    \underset{\alpha}{\sup} \left\vert \left(\widehat{w}^\top U_c - \frac{1}{n_t} \mathbbm{1}_{n_t}^\top U_t\right) \Lambda U^\top \alpha \right\vert,
\end{align*}
where $\widehat{w}$ is the proposed kernel-based estimator obtained via \eqref{eqn:kernel-SBW-QP}.

\begin{theorem} \label{thm:bias-bound}
Suppose that Assumptions (2.1) and (2.2) hold. 
Further assume that $\Vert \widetilde{\bm{K}} - \widetilde{\bm{K}}_l \Vert_2 \leq \frac{\sigma_l - \sigma_{l+1}}{2}$ and $\sup\Vert \alpha \Vert_2 < \infty$. Let $\epsilon \in (0,1)$ be an error parameter, and $W_l$ and $\bm{K}_s$ be the best low-rank approximations for $W^+$ and $\bm{K}$ with ranks $l$ and $s$, respectively. Then, for all the column sampling schemes in Table \ref{tbl:sketch-size-sampling}, with high probability at least $0.9$,  
\begin{equation} \label{eqn:worst-case-bound}
\begin{aligned} 
    \underset{\alpha}{\sup} \left\vert \left(\widehat{w}^\top U_c - \frac{1}{n_t} \mathbbm{1}_{n_t}^\top U_t\right) \Lambda U^\top \alpha \right\vert 
    & = (1+\epsilon)\left\Vert \bm{K} -  \bm{K}_s  \right\Vert_* \\
    & \quad + O\left(\Vert W^{+} - W^{-1}_l \Vert_2 + \Vert W^{+} - W^{-1}_l \Vert_F^2 + \delta\right),
\end{aligned}
\end{equation}
where $\bm{K}_s$ is the best rank-$s$ approximation of $\bm{K}$. $\Vert \cdot \Vert_2$, $\Vert \cdot \Vert_*$ and $\Vert \cdot \Vert_F$ are the 2-norm, the trace norm, and the Frobenius norm, respectively.
\end{theorem}

See Appendix A in Supplementary Materials for the formal definitions of the matrix norms and the proof. 
The above theorem provides an upper bound on the bias and dissects it into three components. The first component, $(1+\epsilon)\left\Vert \bm{K} -  \bm{K}_s  \right\Vert_*$, pertains to the bias due to the degree of kernel matrix approximation. It provides guarantees on relative-error trace norm approximation of the rank-$s$ Nystr\"{o}m approximation without regularization. While the magnitude of this bias component may be reduced by choosing a larger value for $s$, this decision must be informed by choices for the other scaling parameters $l$ and $m$ as well as our error tolerance (see Remark \ref{rmk:scaling-parameter-selection}). The second component of the bias, $ \Vert W^{+} - W^{-1}_l \Vert_2 + \Vert W^{+} - W^{-1}_l \Vert_F^2$, represents the additive error due to the regularization step described in \eqref{eqn:Nystrom-approx-K}, and declines to zero with no regularization. Typically, we expect the second component to be much smaller than the first. The third and final component is the bias from the residual covariate imbalance after weighting, which is directly controlled by the tolerance vector $\delta$ governing the balancing constraints. 

\begin{remark}\label{rmk:scaling-parameter-selection}[Scaling parameter selection] 
    {The upper bound in (3.13) depends on sketching schemes, as well as the scaling parameters $m, l, s$, error parameter $\epsilon$, and matrix coherence $\mu \in \left[1, \frac{n}{s}\right]$ of the dominant $s$-dimensional singular space of $\bm{K}$. Although it is often not straightforward to characterize this upper bound in real datasets, some guidelines for selecting scaling parameters could be established.}
    In Table \ref{tbl:sketch-size-sampling}, we provide sufficient sketch sizes $m$ in terms of $s, \epsilon, \mu$. 
    {While a larger $s$ better approximates the original kernel matrix, it must be some fraction of $m$, and it is desirable to set $m$ as small as possible since it is a proxy for available computational resources. Thus, balancing these two sources of error requires choosing an $s$ value that is sufficiently large to reduce error from matrix approximation, while also being a sufficiently small fraction of $m$. It is known that when uniform sampling is used to form the Nystrom approximation $m=\tilde{O}\left(\frac{s\mu}{\epsilon}\right)$ \citep[][Lemma 10]{wang2019scalable}. Hence, if one sets $\log s \sim \frac{1}{\epsilon}$, then the chosen $m$ should be on the order of $\frac{s\mu}{\epsilon}$. When the budget of column samples $m$ must to be fixed, the target rank could be $s \sim \frac{m}{\mu/\epsilon}$ where $m$ is at least $\mu/\epsilon > 1$ times larger than $s$.} The coherence parameter $\mu$ can be estimated from the truncated SVD of $\bm{K}$ \citep[see, e.g.,][]{mohri2011can}. The regularization parameter $l$ should remain as close to $m$ as possible as long as $\sigma_l$ is not extremely small, guaranteeing numerical stability of $\Lambda_{W,l}^{-1}$. Unfortunately, there is no theory for the choice of $l$ as it is related to the finite precision error. Following \citet{wang2019scalable}, we apply a simple heuristic that $l=\ceil{\frac{s+m}{2}}$. Future theoretical and empirical work is required for optimal scaling parameter selection, as well as characterizing how large the upper bound could be in real datasets.
\end{remark}

%% file: small_sec4.tex
\section{Simulation study}  \label{sec:simulation}
We study the computational performance and estimation
accuracy of several modeling and balancing weighting methods. We extend the simulation design by \citet{hainmueller2012entropy} to encompass nonlinear treatment assignment processes and sample sizes up to a million.

\subsection{Study design} \label{subsec:study-design}
Consider an observational study with continuous and binary covariates generated as follows
\begin{align*}
    \begin{bmatrix}
    X_1 \\ X_2 \\ X_3
    \end{bmatrix}
    \sim
    N \begin{pmatrix}
    \begin{bmatrix}
    0 \\ 0 \\ 0
    \end{bmatrix},
    \begin{bmatrix}
    2 & 1 & -1 \\ 
    1 & 1 & -0.5 \\ 
    -1 & -0.5 & 1
    \end{bmatrix}
    \end{pmatrix},
    \quad X_4 \sim \text{Unif}[-3,3],
    \quad X_5 \sim \chi^2_1,
    \quad X_6 \sim \text{Bern}[0.5].
\end{align*}
Using these six covariates, we generate the outcome and treatment variables as
\begin{align}
    & Y = (X_1 + X_2 + X_5)^2 + \eta, \quad \eta \sim N(0,1),  \label{eqn:outcome-design} \\
    & A = \mathbbm{1}\left\{X_1^2 + 2X_2^2 - 2X_3^2 - (X_4+1)^3 - 0.5\log(X_5+10) + X_6 -1.5 + \varepsilon > 0 \right\}, \label{ps-design:nonlinear}
\end{align}
where $\varepsilon \sim N(0,\sigma^2_{\varepsilon})$. \cmmnt{By adding more nonlinearity, \eqref{ps-design:nonlinear} is extended from the treatment assignment mechanism in \citet{hainmueller2012entropy}.} We consider two settings with $\sigma^2_{\varepsilon}=30$ and $\sigma^2_{\varepsilon}=100$, hereafter referred to as the \textit{weak overlap} and \textit{strong overlap} settings, respectively. 
We define untransformed covariates $X=(X_1,X_2, X_3,X_4,X_5,X_6)$ and transformed covariates $X^*=(X_1X_3, X_2^2, X_4, X_5, X_6)$. 
For each $\sigma^2_{\varepsilon}$, we consider each interaction of treatment assignment mechanism and accuracy of covariate specification across five sample sizes: $n \in \{2\cdot10^3, 5\cdot10^3, 10^4, 10^5, 10^6\}$. In this design, the ATT equals zero by construction. To separate design from analysis and focus on scalability and speed, we based the weighting estimator (\ref{eqn:linear-estimator}) solely on $\widehat{\pi}$; however, it can be augmented by an outcome model to form a doubly-robust estimator as in \cite{robins1995semiparametric}.

Estimation of the ATT by weighting requires explicit modeling of $\pi$ for $\widehat{\psi}_{\text{mod}}$ in the modeling approach, or solving the QP \eqref{eqn:kernel-SBW-QP} to obtain $\widehat{\psi}_{\text{bal}}$ in the balancing approach. We consider various methods, including some considered useful for large-scale modeling or optimization. For practicality, we focus on methods in readily available \texttt{R} packages. For consistent comparisons of balancing estimators, we fix $\delta=0.0005$ across all methods. However, we also evaluate performance when selecting $\delta$ in a data-driven fashion using the algorithm in \cite{chattopadhyay2020balancing}.

We consider nine methods for $\widehat{\psi}_{\text{mod}}$ and seven methods for $\widehat{\psi}_{\text{bal}}$, for a total of 16 methods classified into six groups (see Table \ref{tbl:methods}. For $\widehat{\psi}_{\text{mod}}$, Group M1 methods use logistic regression (or GLM) to model the propensity score. Group M2 contains lasso algorithms with the tuning parameter selected via 10-fold cross validation. In Groups M1 and M2, the regression models for $\pi$ add quadratic terms and all pairwise products of observed covariates. Group M3 includes three common nonparametric algorithms: random forests, kernel ridge regression, and generalized additive models using splines. \cmmnt{When non-parametric methods are employed in the modeling approach, we must rely on either empirical process conditions or cross fitting to estimate $\widehat{\pi}$. Here, we use the two-fold cross fitting in the same spirit of \citep[see, e.g.,][]{Chernozhukov18, kennedy2020sharp}.} For $\widehat{\psi}_{\text{bal}}$, we solve \eqref{eqn:kernel-SBW-QP} where we use the Gaussian kernel and set $m=300$, $l=200$, $s=100$ for the Nystr\"{o}m approximation procedure. Group B1 includes open-source QP solvers in \texttt{R}. Group B2 studies two commercial solvers. Finally, Group B3 includes three state-of-the-art ADMM-based solvers aimed at reducing processing time and used with the kernel balancing approach. For OSQP, we use the same parameters as \citet{osqp2020}. 

We assess estimator performance using the root-mean-squared error (RMSE) defined by $\left\{\frac{1}{S}\sum_{j=1}^S (\widehat{\psi}_j - \psi)^2\right\}^{1/2}$ averaged across $S=500$ simulations, and quantify the computational cost in average central processing unit (CPU) time across simulations, which we report in seconds unless otherwise noted. \cmmnt{Unless otherwise mentioned, each method is implemented in \texttt{R} using each package's internally available solvers and default settings for ease of performance comparisons across methods. However, some algorithms may perform better under non-default settings.} All simulations were completed using the Harvard Faculty of Arts and Sciences Research Computing clusters with available memory fixed at 128GB. In the results tables, we write `F' when the optimization process failed in more than half the simulations, and `M' when the system ran out of memory in more than half the simulations.

 \begin{table}[!t]
\centering
\caption{Description of modeling and solution methods used in the simulation study. 
}
\begin{tabular}{@{}clll@{}}
\toprule
                                                                              & Method                                                                          & R package                            & Description \\ \hline 
\multirow{8}{*}{\begin{tabular}[c]{@{}c@{}}\\ \\ \\  \\ \\ \\ Modeling\\ Approach\\  \end{tabular}}  & \multirow{2}{*}{\begin{tabular}[c]{@{}l@{}} \\ M1. Logistic regression \end{tabular}}                                                & \texttt{glm}        &   \begin{tabular}[c]{@{}l@{}} Commonly used method to fit logistic \\ regression models in R\end{tabular}          \\
                                                                              &                                                                                 & \texttt{sgdglm}     &   \begin{tabular}[c]{@{}l@{}}Stochastic gradient descent methods for \\ estimation with big data \citep{tran2015stochastic}\end{tabular}          \\ \cmidrule(l){2-4} 
                                                                              & \multirow{7}{*}{M2. Regularized regression}                                                       & \texttt{glmnet}     &  \begin{tabular}[c]{@{}l@{}} Lasso via coordinate descent \\ \citep{friedman2010regularization} \end{tabular}           \\
                                                                              &                                                                                 & \texttt{biglasso}   &    \begin{tabular}[c]{@{}l@{}}Scalable lasso for big data\\ \citep{zeng2017biglasso}\end{tabular}        \\
                                                                              &                                                                                 & \texttt{oem}        &   \begin{tabular}[c]{@{}l@{}} Lasso for tall (large n) data \\ \citep{xiong2016orthogonalizing} \end{tabular}          \\
                                                                              &                                                                                 & \texttt{penreg} &  \begin{tabular}[c]{@{}l@{}}ADMM-based lasso implementation\\ \citep{boyd2011distributed} (\texttt{admm.lasso} in results) \end{tabular}           \\ \cmidrule(l){2-4} 
                                                                              & \multirow{4}{*}{M3. Non-parametric  regression}                                              & \texttt{ranger}     &  Fast implementation for Random Forests           \\
                                                                              &                      
                                                                              & \texttt{DRR}     &  
                                                                              \begin{tabular}[c]{@{}l@{}} Fast implementation for Kernel Ridge \\ Regression (\texttt{kernel.ridge} in results) \end{tabular}  \\
                                                                              &                      & \texttt{bam}        &   \begin{tabular}[c]{@{}l@{}}Generalized additive model using splines \\ for big data \citep{wood2015generalized} \end{tabular}          \\ \midrule
\multirow{7}{*}{\begin{tabular}[c]{@{}c@{}} \\  \\ \\ \\ \\ \\ Balancing\\ Approach \end{tabular}} & \multirow{3}{*}{B1. Open-source  solvers}                                                   & \texttt{quadprog}   &  \begin{tabular}[c]{@{}l@{}}Commonly-used QP solver based on the \\dual method  \citep{goldfarb1983numerically}\end{tabular}           \\
                                                                              &                                                                                 & \texttt{qpoases}    &   \begin{tabular}[c]{@{}l@{}}QP solver based on parametric \\ active-set algorithm \citep{ferreau2014qpoases} \end{tabular}        \\ \cmidrule(l){2-4} 
                                                                              & \multirow{3}{*}{\begin{tabular}[c]{@{}l@{}}B2. Commercial  solvers\end{tabular}} & \texttt{gurobi}     &   \begin{tabular}[c]{@{}l@{}}Large scale commercial optimizer \\ by \textit{Gurobi Optimization}\end{tabular}          \\
                                                                              &                                                                                 & \texttt{Rmosek}     &   \begin{tabular}[c]{@{}l@{}}Large scale commercial optimizer \\ by \textit{Mosek ApS} \end{tabular}          \\ \cmidrule(l){2-4} 
                                                                              & \multirow{5}{*}{\begin{tabular}[c]{@{}l@{}}B3. ADMM-based solvers \end{tabular}} & \texttt{osqp}       &   \begin{tabular}[c]{@{}l@{}}Operator splitting solver for QP (OSQP)\\  \citep{osqp2020} \end{tabular}         \\
                                                                              &                                                                                 & \texttt{pogs}       &   \begin{tabular}[c]{@{}l@{}}ADMM-based graph form solver (POGS)\\ \citep{fougner2018parameter} \end{tabular}            \\
                                                                              &                                                                                 & \texttt{scs}        &     \begin{tabular}[c]{@{}l@{}}Operator splitting conic solver (SCS)\\ \citep{o2016conic}  \end{tabular}         \\ \bottomrule
\end{tabular}

\label{tbl:methods}
\end{table}


\subsection{Empirical results} 
\label{subsec:simulation-results}
\cmmnt{
We report empirical results as follows.
Section \ref{subsec:weak_overlap} describes estimation accuracy, CPU time, and convergence for the weak overlap setting. 
Section \ref{subsec:strong_overlap} shows results under strong overlap. Section \ref{subsec:factorization} analyzes the impact of factorization caching and warm starting. Section \ref{subsec:observations} provides general observations. All supporting numerical results are in the appendices.
}
\cmmnt{
\subsubsection{Weak overlap }
\label{subsec:weak_overlap}
}

\begin{table}[!t]
\caption{Root-mean square error (RMSE) under weak overlap. Each cell contains the RMSE for each combination of method, sample size, and covariate specification, averaged across 500 simulations. ``M" signifies the system ran out of memory in more than half the simulations, while ``F" signifies that the optimization process failed to obtain a weighting solution in more than half the simulations.}  \label{tbl:sim-rmse-NL}
\begin{tabular}{@{}clllllllllll@{}}
\toprule
\multicolumn{1}{l}{}       &            & \multicolumn{5}{c}{$X$ correctly specified ($X$)}                                     & \multicolumn{5}{c}{$X$ transformed ($X^*$)}         \\ \midrule 
\multicolumn{2}{l}{\diagbox{Method}{$n$}} &
  \multicolumn{1}{c}{2k} &
  \multicolumn{1}{c}{5k} &
  \multicolumn{1}{c}{10k} &
  \multicolumn{1}{c}{100k} &
  \multicolumn{1}{c}{1M} &
  \multicolumn{1}{c}{2k} &
  \multicolumn{1}{c}{5k} &
  \multicolumn{1}{c}{10k} &
  \multicolumn{1}{c}{100k} &
  \multicolumn{1}{c}{1M} \\ \midrule
\multirow{9}{*}{$\widehat{\psi}_{\text{mod}}$} &
  \texttt{glm} &
  1.72 &
  1.29 &
  1.20 &
  1.19 &
  1.18 &
  1.85 &
  1.37 &
  1.30 &
  1.18 &
  1.18 \\
                           & \texttt{sgdglm}     & 1.72 & 1.29 & 1.20 & 1.19 & 1.18     & 1.86 & 1.37 & 1.31 & 1.18 & 1.19 \\
                           & \texttt{glmnet}     & 2.20 & 1.18 & 0.68 & 0.23 & 0.12   & 4.57 & 1.79 & 0.67 & 0.39 & 0.35 \\
                           & \texttt{biglasso}   & 2.57 & 1.08 & 0.65 & 0.22 & 0.12   & 5.23 & 1.68 & 0.68 & 0.39 & 0.35 \\
                           & \texttt{oem}        & 3.37 & 1.27 & 0.79 & 0.26 & 0.12   & 6.47 & 2.97 & 0.79 & 0.40 & 0.35 \\
                           & \texttt{admm.lasso} & 3.83 & 1.59 & 0.84 & 0.29 & 0.15   & 7.29 & 3.04 & 1.07 & 0.70 & 0.51 \\
                           & \texttt{ranger}     & 0.71 & 0.33 & 0.22 & 0.15 & 0.09   & 0.97 & 0.41 & 0.36 & 0.27 & 0.22 \\
                           & \texttt{kernel.ridge}  & 1.87 & 1.19 & 1.01 & 0.89 & M   & 6.26 & 6.07 & 5.59 & 4.87 & M \\
                           & \texttt{bam}        & 2.84  & 1.31   & 1.20 & 1.19   & 1.18    & 3.37 & 1.95 & 1.25 & 1.19 & 1.18 \\ \midrule
\multirow{7}{*}{$\widehat{\psi}_{\text{bal}}$} & \texttt{quadprog}   & 0.31 & 0.20 & 0.15 & F    & F     & 0.51 & 0.34 & 0.33 & F    & F    \\
                           & \texttt{qpoases}    & 0.32 & 0.22 & 0.15 & F    & F      & 0.65 & 0.34 & 0.24 & F    & F    \\
                           & \texttt{gurobi}     & 0.19 & 0.13 & 0.09 & 0.04 & 0.02   & 0.31 & 0.19 & 0.15 & 0.08 & 0.05 \\
                           & \texttt{Rmosek}      & 0.19 & 0.13 & 0.10 & 0.04 & 0.02    & 0.31 & 0.20 & 0.15 & 0.08 & 0.05 \\
                           & \texttt{osqp}       & 0.19 & 0.03 & 0.10 & 0.04 & 0.02  & 0.31 & 0.19 & 0.16 & 0.08 & 0.05 \\
                           & \texttt{pogs}       & 0.31 & 0.20 & 0.15 & {M}    & {M}     & 0.59 & 0.36 & 0.29 & {M}    & {M}    \\
                           & \texttt{scs}        & 0.25  & 0.22  & 0.19  & 0.15  & 0.10  & 0.55 & 0.38 & 0.25 & 0.18 & 0.15 \\ \bottomrule
\end{tabular}
\end{table}

\cmmnt{\textbf{RMSE.} }Table \ref{tbl:sim-rmse-NL} shows estimation accuracy under weak overlap. The proposed kernel balancing estimators $\widehat{\psi}_{\text{bal}}$ substantially outperform the modeling estimators $\widehat{\psi}_{\text{mod}}$ at all considered sample sizes, especially in small to moderate ones, regardless of covariate specification; however, gaps between $\widehat{\psi}_{\text{bal}}$ and $\widehat{\psi}_{\text{mod}}$ are smaller with transformed covariates $X^*$. Performance varied greatly within the class of modeling approaches; in particular, random forests were most successful at larger sample sizes. \cmmnt{We note that the RMSE for OSQP is the same as or only slightly different from the commercial solvers; however, these commmercial solvers also used the Nystr\"{o}m approximation for improved computational efficiency.} We do not report results for kernel-based stable balancing weights without the Nystr\"{o}m approximation, since no methods were successful for samples of 100,000 or more. 

\cmmnt{\textbf{CPU time.} }Because the average CPU time did not appreciably vary across simulation settings, we present only their averages in Table \ref{tbl:sim-Qvg-timing}. For $\widehat{\psi}_{\text{bal}}$, forming the kernel basis requires less computation time than solving the QP, demonstrating the efficiency of the  Nystr\"{o}m approximation {in the kernel balancing algorithm}.  When solving the same QP \eqref{eqn:kernel-SBW-QP}, OSQP was faster than the commercial solvers MOSEK and Gurobi, by about an order of magnitude {in the case of} the latter. The GLM and OSQP methods were fastest, taking only seconds to find the optimal weights even in the largest sample size of one million. In contrast, the lasso and non-parametric methods required much more time at that sample size, in some cases hundreds of times more than GLM and OSQP. Although not presented here, the OSQP and GLM-based estimators took under a minute on average to handle a simulated dataset of 10 million observations, whereas the slowest methods took up to days. \cmmnt{Among the ADMM-based solvers, OSQP was much faster than SCS at the largest sample size.}

\begin{table}[!t]
\caption{Average CPU time across simulation settings. Each cell contains the CPU time for each combination of method, sample size, and covariate specification, averaged across 500 simulations. All values are reported in seconds unless specified as being in hours (``hr"). ``M" signifies the system ran out of memory in more than half the simulations, while ``F" signifies that the optimization process failed to obtain a weighting solution in more than half the simulations.} \label{tbl:sim-Qvg-timing}
\centering
\begin{tabular}{@{}clrrrrc@{}}
\toprule
\multicolumn{1}{l}{}       &            & \multicolumn{5}{c}{Time (in seconds unless otherwise noted)}                                         \\ \midrule 
\multicolumn{2}{l}{\diagbox{Method}{$n$}} &
  \multicolumn{1}{c}{2k} &
  \multicolumn{1}{c}{5k} &
  \multicolumn{1}{c}{10k} &
  \multicolumn{1}{c}{100k} &
  \multicolumn{1}{c}{1M} 
  \\ \midrule
\multirow{9}{*}{$\widehat{\psi}_{\text{mod}}$} &
  \texttt{glm} &
  \textless{}0.1 &
  \textless{}0.1 &
  \textless{}0.1 &
  0.4 &
  8.6  \\
                           & \texttt{sgdglm}    & \textless{}0.1 & \textless{}0.1 & \textless{}0.1 & 0.3 & 5.0      \\
                           & \texttt{glmnet}     & 1.0            & 1.8            & 4.4            & 18  & 362    \\
                           & \texttt{biglasso}   & 9.2            & 12            & 20            & 266  & 0.8hr  \\
                           & \texttt{oem}        & 38            & 81             & 224             & 0.6hr & 5hr    \\
                           & \texttt{admm.lasso} & 1.3            & 2.7            & 6.3            & 15  & 311    \\
                           & \texttt{ranger}     & 8.7            & 30            & 128            & 0.4hr  & 3hr  \\
                           & \texttt{kernel.ridge} & 1.1            & 3.6            & 29            & 0.3hr  & M  \\
                           & \texttt{bam}        & 0.7            & 1.6            & 3.8            & 21  & 210   \\ \midrule
\multirow{7}{*}{$\widehat{\psi}_{\text{bal}}$} & \texttt{quadprog}   & 0.9            & 15             & 119            & F    & F        \\
                           & \texttt{qpoases}    & 48             & 127            & 392            & F    & F       \\
                           & \texttt{gurobi}     & \textless{}0.1 & \textless{}0.1 & 0.5 & 6.3  &  94    \\
                           & \texttt{Rmosek}      & \textless{}0.1 & \textless{}0.1 & \textless{}0.1            & 4.2   & 37     \\
                           & \texttt{osqp}       & \textless{}0.1 & \textless{}0.1 & \textless{}0.1 & 3.2 & 9.6    \\
                           & \texttt{pogs}       & 3.7            & 46             & 305            & M   & M         \\
                           & \texttt{scs}        &  \textless{}0.1  &  5.3          &  14.5         & 27    & 169   \\ \bottomrule
\end{tabular}
\end{table}

\cmmnt{\textbf{Convergence and Memory.} }Several methods faced difficulties at larger sample sizes, particularly $\widehat{\psi}_{\text{bal}}$ approaches. At samples $\geq$ 100,000, the open-source solvers failed to converge, while the system ran out of memory when attempting to execute POGS, perhaps because it cannot accommodate sparse matrices. Among $\widehat{\psi}_{\text{mod}}$ approaches, kernel ridge regression ran out of memory at 1 million observations.

\cmmnt{
\subsubsection{Strong overlap }
\label{subsec:strong_overlap}
}

Results for RMSE, CPU time, convergence, and memory exhaustion in the strong covariate overlap setting were generally consistent with those in weak overlap, but with smaller performance gaps between $\widehat{\psi}_{\text{mod}}$ and $\widehat{\psi}_{\text{bal}}$ (see Appendix C for details). 


\cmmnt{
\subsubsection{Factorization caching and warm starting }
\label{subsec:factorization}
}
As noted in Section \ref{subsec:fast-optimization}, the OSQP algorithm facilitates factorization caching and warm starting for even better computational efficiency when multiple iterations are required, such as for data-driven selection of $\delta$. To demonstrate, we repeatedly solved a sequence of OSQP-based weighting problems in the same simulation settings using $100$ different values of $\delta$. The results showed that the OSQP algorithm achieved an additional $2.4$-fold improvement in CPU time on average, whereas the commercial solvers did not exhibit such efficiency gains (see Appendix D).

%% file: small_sec5.tex
\section{A national study of heart attack care by hospital profit status}

Heart attacks are generally treated with either interventional cardiology through percutaneous coronary intervention (PCI), or alternatively medical management. In the more severe subtype known as ST-elevation MIs, PCI is generally recommended. However, older patients are much more likely to have non-ST-elevation MIs, and guidelines are less clear on PCI use in this subtype and more generally among older patients with complex comorbid conditions \citep{acc2021pci}. Thus, clinical decision-making in this setting can be affected by both individual physician practice style and the norms of the institutions where they work. Medicare reimburses hospitals more when patients receive interventional cardiology, possibly incentivizing for-profit hospitals to do so more often. We asked whether for-profit hospitals were more likely to use interventional cardiology on heart attack patients than other hospitals, and whether patients outcomes. differed Two previous studies reported opposing results on relationships between profit status, PCI use, and outcomes. \cite{sloan2003does} found for-profit hospitals had higher rates of PCI than other hospitals yet similar mortality rates. In contrast, \cite{shah2007impact} found similar rates of PCI use among non-ST elevation MI cases. However, these studies had fewer than 160,000 patients and used data from small numbers of hospitals in voluntary reporting systems; for example, \cite{shah2007impact} studied only 58 for-profit hospitals. 

To build on previous work by applying our proposed methods to a much larger and more representative sample, we studied a national dataset of Medicare administrative claims from the Centers for Medicare \& Medicaid Services (CMS) for patients admitted to short-term acute-care hospitals for initial treatment of heart attack. We studied 1,273,636 Medicare beneficiaries hospitalized for heart attack between October 1, 2010 and November 30, 2019. We defined treatment as admission to a for-profit hospital, and control as admission to other types of hospitals. 

A fair comparison of for-profit and control hospitals requires careful adjustment for differences in potentially confounding covariates. For example, heart attack treatment and outcomes can be affected by patient demographics and contemporary clinical guidelines. Thus, we adjusted for patient age on admission, sex, race, and admission year. The type of heart attack also strongly affects treatment and prognosis, as ST-elevation Myocardial Infarctions (STEMIs) are much more likely to receive PCI but are also much more severe than Non-ST-Elevation Myocardial Infarctions (NSTEMIs). Thus, we defined these by checking the inpatient record for corresponding International Classification of Diseases, 9th Revision, Clinical Modifictation (ICD-9-CM) or ICD-10-CM principal diagnosis codes. We then included covariates for these types of heart attack in the adjustment model. We also adjusted for emergent admission type and history of PCI or coronary artery bypass grafting (CABG) surgery. As comorbid risk factors such as history of diabetes, renal disease, or stroke can affect both prognosis and likelihood of receiving PCI, we also checked the inpatient records for evidence of a set of 43 comorbidities drawn from risk models maintained by CMS, concentrating on conditions noted as present on admission (POA). This screening only counted diagnoses explicitly documented as already existing upon the patient's admission to the hospital rather than those noted as having occurred after the patient was admitted.

Patients may choose where to live in part on local healthcare resources, which can affect both selection into hospitals and the outcome. Thus, we also defined county-level measures of hospital characteristics such as academic affiliation, nurse staffing levels, size in number of beds, available cardiac technologies, and urban location, for a total of 17 hospital characteristics. This partially addresses selection bias due to 
textcolor{red}{residential choice} based on potential medical need. 

While the simulations included 6 adjustment covariates, our case study includes 70 patient covariates and 17 area-level hospital characteristics for a total of 87 covariates. Given the additional complexity arising from this much number of covariates and the size of the dataset (over 1 million control patients and 1.27 million in total), we proceeded with the stable kernel balancing approach as it showed excellent computational efficiency without diminished estimator accuracy.

To make the assumption of conditional exchangeability more plausible in this demonstration of the method, we were careful to leverage the Medicare data as described above to adjust for as many observable pre-treatment covariates as possible such as demographics, heart attack type, risk factors, and hospital market characteristics. However, as this is an observational study, it is possible for there to be omitted unobserved confounders.

We used the proposed weighting methods to balance the 87 covariates and functions thereof, in order to estimate the average treatment effect on the treated; i.e., the effect of for-profit admission on the likelihood of receiving PCI. We assessed balance using target absolute standardized mean differences (TASMD) \citep{chattopadhyay2020balancing}, considering the for-profit and control groups balanced if no TASMD exceeded 0.1. Because the same set of weights can in principle be used to examine multiple endpoints, we used the weights to assess PCI use and several outcomes: mortality within 30 days of admission, length of stay, and readmission within 30 days of discharge.


\subsection{Kernel balance}

Table \ref{tbl:sample} shows selected characteristics of 206,948 patients from 775 for-profit and 1,066,780 patients from 2,763 control hospitals. See Appendix E for a balance table showing all 87 covariates.

\begin{table}[t!]
\caption{Selected Covariates and Sample Size before Weighting. The table shows selected unweighted covariate means and sample sizes among all patients admitted to for-profit hospitals or other hospitals, along with the target absolute standardized mean difference (TASMD) before weighting.} \label{tbl:sample}
\centering
\begin{tabular}{lrrr}
  \hline
 Covariate (\% unless noted) / Sample Size   & For-Profit & Controls & TASMD Before \\ 
  \hline
  \emph{Demographic Characteristics} & & & \\
  \hspace{.5cm} Age at admission (years, mean) & 79.0 & 79.3 & 0.03 \\ 
  \hspace{.5cm} Female & 47.8\% & 47.7\% & 0.00 \\ 
  \hspace{.5cm} Non-Hispanic White & 79.0\% & 84.2\% & 0.13 \\ 
  \hspace{.5cm} Black & 8.3\% & 7.8\% & 0.02 \\ 
  \hspace{.5cm} Hispanic & 8.6\% & 4.4\% & 0.15 \\ 
  \hline
  \emph{Comorbidities Present on Admission} & & & \\
  \hspace{.5cm} Diabetes & 37.6\% & 37.1\% & 0.01 \\ 
  \hspace{.5cm} Heart Failure & 43.5\% & 43.8\% & 0.01 \\ 
  \hspace{.5cm} Valvular Disease & 15.6\% & 19.1\% & 0.09 \\ 
  \hline
  \emph{County of Residence Hospital Characteristics} & & & \\
  \hspace{.5cm} Small Size ($<$ 251 beds) & 79.5\% & 72.9\% & 0.44 \\ 
  \hspace{.5cm} Nonteaching  & 87.6\% & 81.5\% & 0.41 \\ 
   \hline 
   \emph{Sample Size} & & & \\
  \hspace{.5cm} Patients & 206,948 & 1,066,670 &  \\ 
  \hspace{.5cm} Hospitals & 775 & 2,763 &  \\ 
  \hline
\end{tabular}
\end{table}

Because heart attack is an acute condition, the time and place of its occurrence is not easily predictable and patients are usually brought to the closest hospital; hence, many characteristics such as age and several comorbidities are fairly well-balanced, although a number of imbalances persist and require adjustment. While the occurrence of heart attack in relation to the type of nearby hospital may be somewhat haphazard, the distribution of patient race as well as the characteristics of the hospitals in their counties are not. In particular, patients admitted to for-profit hospitals were more likely to be nonwhite, and more likely to reside in counties where hospitals tended to be smaller and nonteaching. Given the policy importance of health equity, these community differences emphasize the need to benchmark for-profit hospital performance.

We used the same kernel and scaling parameters as in the simulations. However, to prioritize estimator accuracy in the case study, we used the high-accuracy setting with a much stricter termination criterion and solution polishing \citep[][Section 4]{osqp2020}, and set $\delta=0$ rather than $\delta=0.0005$. These made the QP more difficult to solve, but favored estimator accuracy. 

\begin{figure}
    \centering
    \includegraphics[scale=0.5]{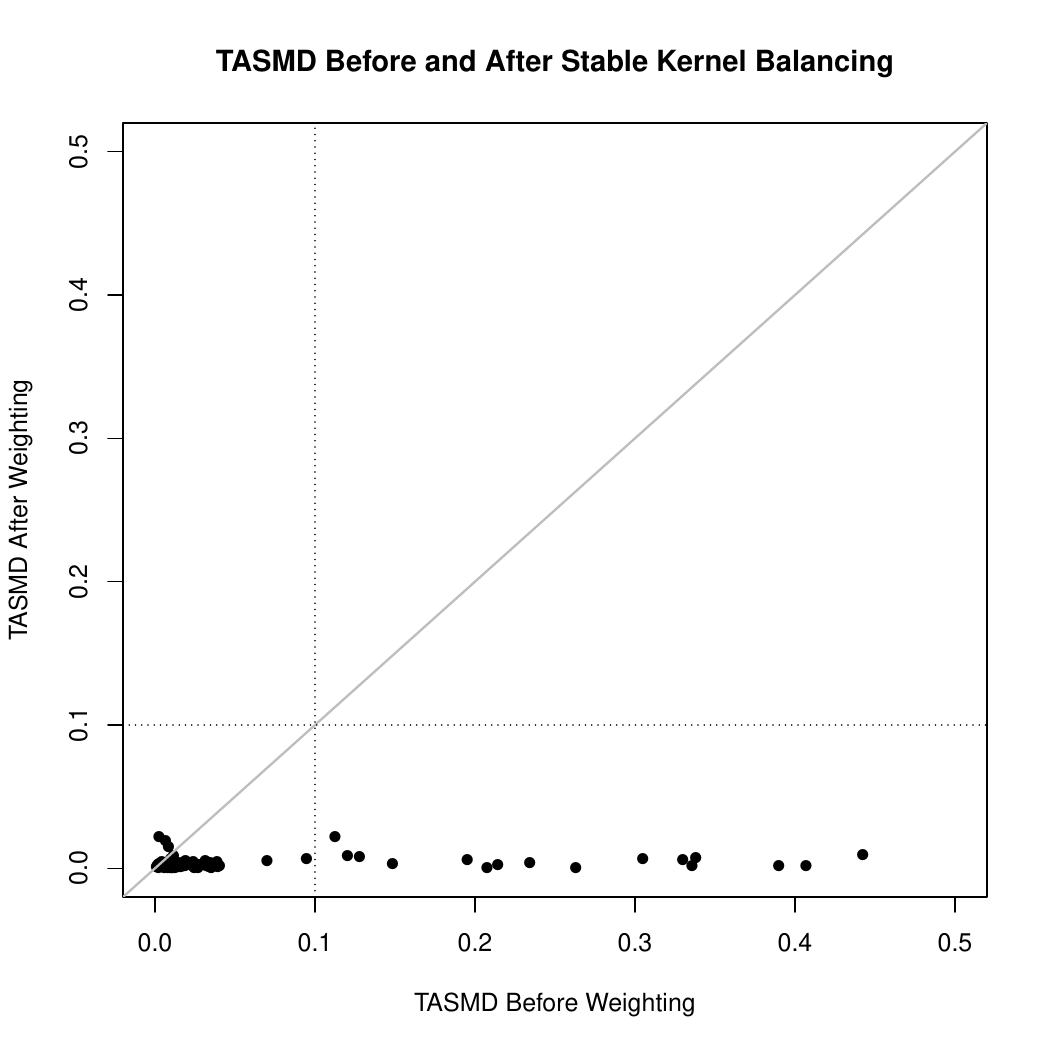}
    \caption{Each dot represents the {Target Absolute Standardized Mean Difference (}TASMD{)} of one of 87 covariates between for-profit and control hospitals before and after weighting.}
    \label{fig:tasmd}
\end{figure}

The stable kernel balancing algorithm efficiently produced excellent balance. Figure \ref{fig:tasmd} shows that despite large initial imbalances, the weighting algorithm successfully balanced all covariates with all TASMDs after weighting well below 0.1, and none exceeding 0.03. 
In our dataset of 1.27 million patients, the kernel basis was computed in just 32.3 seconds, while the OSQP algorithm solved for the stable kernel balancing weights in 132 seconds. 

\subsection{Heart attack treatment and outcomes}
Table \ref{tbl:fp_outcomes} shows treatment and outcomes by hospital profit status after weighting. We found that for-profit hospitals used PCI at almost the same rate as control hospitals, which aligns with the more recent {work by} \cite{shah2007impact}, but conflicts with the {earlier} study by \citet{sloan2003does}. Differences between our study and previous ones may also arise because we used a national sample of hospitals rather than {sets} of hospitals in voluntary reporting programs.

Despite similar PCI use, for-profit hospitals had significantly higher 30-day mortality relative to control hospitals, a finding that differs from the older study by \citet{sloan2003does}. Compared to more contemporary work by \citet{horwitz2017hospital}, we found a larger difference in mortality; however, their analysis pooled patients hospitalized for various reasons. Length of stay was significantly but not meaningfully longer at for-profit hospitals. However, readmission rates were significantly and meaningfully higher at for-profit hospitals.


\begin{table}[t!]
\caption{PCI Use and Outcomes by Hospital Profit Status. The table shows outcome means among for-profit hospital patients, weighted outcome means among control hospital patients, and a 95\% confidence interval for the difference.} \label{tbl:fp_outcomes}
\centering
\begin{tabular}{lrrrc}
  \hline
 Measure (percent unless noted) & For-Profit & Control & Difference & 95\% CI \\ 
  \hline
Received PCI & 33.2 & 33.1 & 0.1 & (-0.1, 0.3) \\ 
30-day Mortality & 14.3 & 13.9 & 0.4 & (0.2, 0.6) \\ 
Length of Stay (days) & 5.76 & 5.63 & 0.13 & (0.10, 0.15) \\ 
30-day Readmission/IP Death & 19.2 & 18.2 & 0.9 & (0.7, 1.1) \\ 
30-day Readmission Only & 11.5 & 10.6 & 0.8 & (0.6, 0.9) \\ 
   \hline
\end{tabular}
\end{table}


%% file: small_sec6.tex
\section{Summary}

In this paper, we describe a weighting approach that overcomes previous barriers to the use of the balancing approach at large scales. We show that the rank-restricted Nystr\"{o}m approximation makes it computationally feasible to find a set of general functions of the covariates in an RKHS. We describe how the kernel basis from the Nystr\"{o}m approximation can be efficiently balanced via a modern ADMM-based convex optimization method. Our extensive simulation study shows that the proposed estimator has superior computation speed relative to other balancing approaches and even logistic regression-based modeling approaches. Moreover, this efficiency gain does not diminish the proposed estimator's accuracy, as it performed as well as or better than other approaches in most simulation settings. Hence, our study provides new insights into the performance of weighting estimators in large datasets.

We used the proposed approach in a large observational study of heart attack treatment and outcomes in for-profit and control hospitals in a national Medicare dataset.  We achieved excellent balance in under 3 minutes. After adjustment, for-profit hospitals used PCI at similar rates as others, but had significantly higher readmission rates. Future work could examine the impact of Medicare's Hospital Readmissions Reduction Program on for-profit hospitals. 

%% file: small_sec7.tex

\section{Software}
The proposed algorithm can be implemented in \texttt{R} with the code provided in \url{https://github.com/kwangho-joshua-kim/osqp-kernel-sbw}.

%% file: small_supp.tex

\begin{center}
{\bf \large Supplementary Material for ``Scalable kernel balancing weights in a nationwide observational study of hospital profit status and heart attack outcomes" by Kwangho Kim, 
Bijan A. Niknam, and 
José R. Zubizarreta}
\end{center}

\section*{Appendix A. Proof of Theorem 1} \label{appsec:proof-thm-1}
We use three matrix norms for our result: for a matrix $A$ with the $(i,j)$-entry $a_{ij}$, we define
\begin{align*}
    \text{Frobenius Norm:}\quad & \Vert A \Vert_F = \sqrt{\sum_{i,j}a_{ij}^2} = \sqrt{\sum_i \sigma_i^2(A)} \\
    \text{Spectral Norm:}\quad & \Vert A \Vert_2 = \underset{\Vert x \Vert_2=1}{\max} \Vert Ax \Vert_2 = \sigma_1(A) \\
    \text{Trace Norm:}\quad & \Vert A \Vert_* = \sum_i \sigma_i(A).
\end{align*}

\begin{proof}

Recall that from Section \ref{subsec:kernel-basis}, 
\begin{gather*}
    \widetilde{\bm{K}} = CW^{+}C^\top, \\
    \widetilde{\bm{K}}_l = CW^{-1}_lC^\top, \\
    (\widetilde{\bm{K}}_l)_s = DD^\top \quad \text{where} \quad D = C U_{W,l} \Lambda_{W,l}^{-1/2}\tilde{V}_{R,s}.
\end{gather*}
Let $w_t = \frac{1}{n_t} \mathbb{1}_{n_t}$. Then by the Cauchy–Schwarz and triangle inequalities, we have that
\begin{align*}
    & \left\vert \left(\widehat{w}_c^\top U_c - w_t^\top U_t\right) \Lambda U^\top \alpha \right\vert \\
    & \leq \left\vert \left(\widehat{w}_c^\top U_c - w_t^\top U_t\right) \Lambda U^\top \alpha - \left(\widehat{w}_c^\top D_c - w_t^\top D_t\right)D^\top \alpha + \left(\widehat{w}_c^\top D_c - w_t^\top D_t\right)D^\top \alpha  \right\vert \\
    & \leq \Vert \widehat{w}_c \Vert_2 \left\Vert U_c\Lambda U^\top - D_cD^\top  \right\Vert_2 \Vert \alpha \Vert_2 + \Vert w_t \Vert_2\left\Vert U_t\Lambda U^\top - D_tD^\top  \right\Vert_2 \Vert \alpha \Vert_2 \\
    & \quad + \left\vert  \left(\widehat{w}_c^\top D_c - w_t^\top D_t\right)D^\top \alpha  \right\vert \\
    & \leq 2\left\Vert \bm{K} - (\widetilde{\bm{K}}_l)_s  \right\Vert_2  \Vert \alpha \Vert_2 + \left\vert  \left(\widehat{w}_c^\top D_c - w_t^\top D_t\right)D^\top \alpha  \right\vert \\
    & \leq 2 \left\{\left\Vert \bm{K} - (\widetilde{\bm{K}})_s  \right\Vert_2 + \left\Vert (\widetilde{\bm{K}})_s - (\widetilde{\bm{K}}_l)_s  \right\Vert_2\right\}  \Vert \alpha \Vert_2 + \left\vert  \left(\widehat{w}_c^\top D_c - w_t^\top D_t\right)D^\top \alpha  \right\vert.
\end{align*}

When $\Vert \widetilde{\bm{K}} - \widetilde{\bm{K}}_l \Vert_2 \leq \frac{\sigma_l - \sigma_{l+1}}{2}$, by \citet[][Theorem 1]{vu2021perturbation}, we have the following first-order approximation
\begin{align*}
    \Vert (\widetilde{\bm{K}})_s - (\widetilde{\bm{K}}_l)_s \Vert_2 & \lesssim \Vert \widetilde{\bm{K}} - \widetilde{\bm{K}}_l \Vert_2 + \sum_{i=1}^l\sum_{j=l+1}^n\left( \frac{\sigma_j^2}{\sigma_i^2 - \sigma_j^2} + \frac{\sigma_i\sigma_j}{\sigma_i^2 - \sigma_j^2} \right) \Vert \widetilde{\bm{K}} - \widetilde{\bm{K}}_l \Vert_2 + \Vert \widetilde{\bm{K}} - \widetilde{\bm{K}}_l \Vert_F^2 \\
    & \lesssim \Vert W^{+} - W^{-1}_l \Vert_2 + \Vert W^{+} - W^{-1}_l \Vert_F^2.
\end{align*}

Also, by Theorem 1 of \citet{wang2019scalable}, we have
\begin{align*}
    \left\Vert \bm{K} - (\widetilde{\bm{K}})_s  \right\Vert_2 & \leq \left\Vert \bm{K} - (\widetilde{\bm{K}})_s  \right\Vert_* \\ 
    & \leq (1+\epsilon)\left\Vert \bm{K} -  \bm{K}_s  \right\Vert_*
\end{align*}
with high probability at least $0.9$.

For the last term in the last display, by the Cauchy–Schwarz inequality it follows that
\begin{align*}
    \underset{\alpha}{\sup} \left\vert  \left(\widehat{w}_c^\top D_c - w_t^\top D_t\right)D^\top \alpha  \right\vert & \leq \left\Vert \widehat{w}_c^\top D_c - w_t^\top D_t \right\Vert_2 \underset{\alpha}{\sup} \left\Vert D^\top \alpha \right\Vert_2 \\
    & \lesssim \left\Vert \widehat{w}_c^\top D_c - w_t^\top D_t \right\Vert_1 \\
    & \lesssim \delta.
\end{align*}

Hence, provided that $\underset{\alpha}{\sup} \Vert \alpha \Vert_2 < \infty$,
putting the two pieces together we have the desired probabilistic bound
\begin{align*}
    & \underset{\alpha}{\sup} \left\vert \left(\widehat{w}_c^\top U_c - w_t^\top U_t\right) \Lambda U^\top \alpha \right\vert \\
    & = (1+\epsilon)\left\Vert \bm{K} -  \bm{K}_s  \right\Vert_* + O\left(\Vert W^{+} - W^{-1}_l \Vert_2 + \Vert W^{+} - W^{-1}_l \Vert_F^2 + \delta \right).
\end{align*}

\end{proof}



\clearpage

\section*{Appendix B. More details on ADMM and OSQP} \label{appsec:admm-osqp-details}

Convex QPs such as \eqref{eqn:stable-balancing-weights} are well-studied and various solution methods exist. Some commonly used solution methods include {active-set}, {interior-point}, and operator splitting methods.  {Active-set} methods are the traditional algorithms for solving QPs. These algorithms explore the feasible region containing the solution by iteratively adding and dropping active constraints until the solution is found. {Interior-point} methods, the default algorithms used by many modern commercial solvers, solve unconstrained optimization problems for varying barrier functions until the optimum is achieved. 

Typically, active-set methods outperform interior-point methods when the number of covariates or constraints is small, or when warm starts are used. Otherwise, interior-point methods generally outperform active-set methods in terms of scalability and speed. That said, in general the iteration costs of interior-point methods grow non-linearly with the dimensionality of the problem, rendering them impractical for very large-scale settings. 
Alternatively, operator splitting methods obtain an optimal solution using only first-order information about the cost function (i.e., the subgradients). They reduce the optimization problem to that of finding zeros of the sum of monotone operators. Here, we focus on a particular class of operator splitting methods known as the {alternating direction method of multipliers} (ADMM) \citep{boyd2011distributed}) which has been shown to extend very well to large-scale convex optimization problems.

\textbf{ADMM.}
The \emph{Alternating direction method of multipliers} is likely the most commonly used operator splitting method due to its simplicity and ability to accommodate large problems. ADMM solves problems of the form
\begin{equation} \label{eqn:ADMM-form}
\begin{aligned}
    & {\text{minimize}} \quad g(\tilde{v}) + h(v)\\
    & \text{subject to} \quad G\tilde{v} + Hv = q 
\end{aligned}    
\end{equation} 
for two sets of variables $\tilde{v}$ and $v$. Following \citet{osqp2020}, we introduce auxiliary variables $\tilde{w}$ and $\tilde{z}$ and write Problem \eqref{eqn:kernel-SBW-QP} as
\begin{equation} \label{eqn:SBW-ADMM-form}
\begin{aligned}
    & {\text{minimize}} \quad \tilde{w}^\top \mathbb{I}_{n_c} \tilde{w}  - \frac{1}{n_c}\tilde{w}^\top \mathbb{1}_{n_c} + \mathcal{I}_{Qw=z}(\tilde{w}, \tilde{z}) + \mathcal{I}_{\mathcal{C}(\delta)}(z) \\
    & \text{subject to} \quad \mathbb{I}_{1+2n_c+Kd}[\tilde{w}, \tilde{z}]^\top - \mathbb{I}_{1+2n_c+Kd}[w,z]^\top=0,
\end{aligned}    
\end{equation}
where $\mathcal{I}_{S}$ is the indicator function for the set $S$. Problem \eqref{eqn:SBW-ADMM-form} is in the ADMM form given by \eqref{eqn:ADMM-form}, where $g(\tilde{w}, \tilde{z})=\tilde{w}^\top \mathbb{I}_{n_c} \tilde{w}  - \frac{1}{n_c}\tilde{w}^\top \mathbb{1}_{n_c} + \mathcal{I}_{Qw=z}(\tilde{w}, \tilde{z})$ and $h(w,z)=\mathcal{I}_{\mathcal{C}(\delta)}(z)$. Problem \eqref{eqn:SBW-ADMM-form} can be solved via the following ADMM iterations 
\begin{align}
    & (w^{k+1}, \tilde{z}^{k+1}) \leftarrow \underset{Qw=z}{\argmin} \quad {w}^\top \mathbb{I}_{n_c} {w}  - \frac{1}{n_c}{w}^\top \mathbb{1}_{n_c} + \frac{\sigma}{2}\left\Vert x - w^k \right\Vert^2 + \frac{\rho}{2}\left\Vert z - z^k + \frac{y^k}{\rho} \right\Vert^2 \label{eqn:ADMM-iter-1} \\
    & z^{k+1} \leftarrow \Pi_{\mathcal{C}(\delta)}\left(\tilde{z}^{k+1} + \frac{y^k}{\rho} \right) \label{eqn:ADMM-iter-2}\\
    & y^{k+1} \leftarrow y^k + \rho\left(\tilde{z}^{k+1}-z^{k+1}\right) \label{eqn:ADMM-iter-3}
\end{align}
where $y$ is a dual covariate, $\sigma, \rho > 0$ are step-size parameters, and $\Pi_{\mathcal{C}(\delta)}$ denotes the Euclidean projection onto $\mathcal{C}(\delta)$ (see \citealt{boyd2011distributed} for details). 

ADMM blends the advantages of dual decomposition and the method of multipliers (or the augmented
Lagrangian method). First, as a method of multipliers, ADMM exhibits excellent convergence properties. In fact, Problem \eqref{eqn:ADMM-form} does not require any assumptions for convergence beyond convexity of $g$ and $h$ and that its augmented Lagrangian has a saddle point, an advantage not shared by other operator splitting methods.
Second, as described by the steps in equations \eqref{eqn:ADMM-iter-1} and \eqref{eqn:ADMM-iter-2}, in ADMM $w^k$ and $z^k$ are updated sequentially or alternatingly, where each step is typically far less expansive than the original minimization problem jointly over $(w,z)$. Indeed, the inner QP \eqref{eqn:ADMM-iter-1} is an equality-constrained problem which can be reduced to solving a linear system, and \eqref{eqn:ADMM-iter-2} is simply a box projection. This decomposability therefore substantially reduces the computational burden required for execution, and allows ADMM to find solutions much faster than conventional methods.

Despite these appealing features, ADMM has some limitations. One potential limitation is its inability to detect infeasibility. Moreover, it loses its speed advantage when aiming to find high-accuracy solutions, and its convergence rates also highly depend on the parameter selection and the data setting. Despite these challenges, significant progress has recently been made in adapting ADMM to very large-scale convex optimization problems. \citet{o2016conic} and \citet{fougner2018parameter} proposed the splitting-based conic solver (SCS) and proximal operator graph-form solver (POGS), respectively. Based on their own parameter selection heuristics and data preconditioning, they showed that both SCS and POGS scale well to extremely large problems, with SCS also able to detect infeasibility. More recently, \citet{osqp2020} developed the operator splitting solver for quadratic programs (OSQP), a state-of-the-art ADMM-based solver particularly tailored to general QPs. OSQP is the first generic QP solver that addresses all three challenges of detecting infeasibility, finding high-accuracy solutions, and adapting to different data settings.

\begin{algorithm}[t]
\caption{OSQP for SBW}
\label{algo:osqp}
\Input{$Q, l(\delta), u(\delta)$}
\Parameter{$\rho > 0, \sigma > 0, \upalpha \in (0,2)$}
Initialize $w^0, y^0, z^0, \nu^0$ and set $k=0$\\
\Repeat{termination criterion satisfied}{
$(\widetilde{w}^{k+1}, {\nu}^{k+1}) \leftarrow$ solve 
$\begin{bmatrix}
-(1+\sigma)\mathbb{I}_{n_c} & -Q^\top  \\
-Q & \rho^{-1}\mathbb{I}
\end{bmatrix}
\begin{bmatrix}
\widetilde{w}^{k+1}   \\
\nu^{k+1} 
\end{bmatrix}
=
-\begin{bmatrix}
\sigma w^{k} - 1/{n_c}\mathbbm{1}_{n_c}   \\
z^{k} - \rho^{-1}y^k
\end{bmatrix}$ \;
$\widetilde{z}^{k+1} \leftarrow z^k + \rho^{-1}(\nu^{k+1} - y^k)$ \\
$w^{k+1} \leftarrow \upalpha \widetilde{w}^{k+1} + (1-\upalpha)w^k$ \\
$z^{k+1} \leftarrow \Pi_{\mathcal{C}(\delta)}\left(\upalpha\widetilde{z}^{k+1} + (1-\upalpha)z^k + \rho^{-1}y^k \right)$ \\
$y^{k+1} \leftarrow y^k + \rho\left(\upalpha \widetilde{z}^{k+1} + (1-\upalpha)z^k - z^{k+1} \right)$\\
$k \leftarrow k+1$
}
\end{algorithm}

\textbf{OSQP for SBW.}
For convenience, the OSQP algorithm for SBW is presented in Algorithm \ref{algo:osqp} separately from Figure \ref{algo:kernel-osqp}. Steps 3 and 4 correspond to the reduced Karush-Kuhn-Tucker (KKT) system for the inner QP \eqref{eqn:ADMM-iter-1}, and Steps 6 and 7 correspond to \eqref{eqn:ADMM-iter-2} and \eqref{eqn:ADMM-iter-3}, respectively. In Steps 5, 6, and 7, the relaxation method is used for the \emph{w}-, \emph{z}-, and \emph{y}-updates with the relaxation parameter $\upalpha$. The termination criterion is satisfied when both the primal and dual residuals are smaller than some prespecified tolerance, or if infeasibility is detected with some prespecified level of accuracy. Although the method of choosing the optimal parameters $(\rho , \sigma , \upalpha)$ may vary, \citet{osqp2020} provides guidance for doing so with OSQP. The parameter $\sigma$ behaves like a regularization term and can be as small as possible, so long as it ensures that a unique solution of the linear system in Step 3 of Algorithm \ref{algo:osqp} exists. Also, through extensive empirical research, it has been shown that setting the relaxation parameter $\upalpha$ in the range $[1.5,1.8]$ generally improves the convergence rate. To set the step size $\rho$, OSQP adopts heuristics that assign higher values for active constraints and lower values for inactive constraints. Furthermore, to ensure fast convergence over a wide range of problems, the algorithm uses the ratio between the primal and dual residuals to adaptively update $\rho$ from one iteration to the next. \citet{osqp2020} showed that the proposed parameter selection heuristics in OSQP exhibit solid improvements in terms of convergence and stability.

We remark that POGS and SCS are alternative ADMM-based methods that could be applied; however, these approaches require reformulation of our QP, which introduces many extra coefficients into the optimization procedure and therefore may not be computationally efficient \citep{osqp2020}.

\clearpage
\newpage

\section*{Appendix C. RMSE under strong overlap} \label{appsec:strong_overlap}

Results for under strong overlap are presented in Web Table \ref{tbl:sim-rmse-NL-so}.

\begin{table}[h!]
\begin{tabular}{@{}clllllllllll@{}}
\toprule
\multicolumn{1}{l}{}       &            & \multicolumn{5}{c}{$X$ correctly specified ($X$)}                                     & \multicolumn{5}{c}{$X$ transformed ($X^*$)}         \\ \midrule 
\multicolumn{2}{l}{\diagbox{Method}{$n$}} &
  \multicolumn{1}{c}{2k} &
  \multicolumn{1}{c}{5k} &
  \multicolumn{1}{c}{10k} &
  \multicolumn{1}{c}{100k} &
  \multicolumn{1}{c}{1M} &
  \multicolumn{1}{c}{2k} &
  \multicolumn{1}{c}{5k} &
  \multicolumn{1}{c}{10k} &
  \multicolumn{1}{c}{100k} &
  \multicolumn{1}{c}{1M} \\ \midrule
\multirow{9}{*}{$\widehat{\psi}_{\text{mod}}$} &
  \texttt{glm} &
  1.13 &
  0.72 &
  0.48 &
  0.40 &
  0.40 &
  1.31 &
  0.80 &
  0.56 &
  0.41 &
  0.40 \\
                           & \texttt{sgdglm}     & 1.13 & 0.72 & 0.48 & 0.40 & 0.40     & 1.31 & 0.80 & 0.56 & 0.41 & 0.40 \\
                           & \texttt{glmnet}     & 1.10 & 0.87 & 0.58 & 0.17 & 0.07   & 1.30 & 1.23 & 0.51 & 0.21 & 0.13 \\
                           & \texttt{biglasso}   & 1.25 & 0.82 & 0.55 & 0.16 & 0.08   & 1.30 & 1.02 & 0.50 & 0.21 & 0.12 \\
                           & \texttt{oem}        & 1.23 & 0.82 & 0.58 & 0.17 & 0.08   & 1.33 & 1.49 & 0.71 & 0.22 & 0.13 \\
                           & \texttt{admm.lasso} & 1.78 & 0.87 & 0.58 & 0.22 & 0.14   & 2.23 & 1.12 & 0.56 & 0.27 & 0.16 \\
                           & \texttt{ranger}     & 1.22 & 0.86 & 0.66 & 0.22 & 0.09   & 1.27 & 0.85 & 0.52 & 0.20 & 0.07 \\
                           & \texttt{kernel.ridge}     & 1.18 & 0.84 & 0.67 & 0.29 & M   & 1.53 & 0.91 & 0.67 & 0.32 & M \\
                           & \texttt{bam}        & 2.47 & 0.67 & 0.46 & 0.41 & 0.40    & 3.37 & 1.95 & 1.25 & 1.19 & 1.18 \\ \midrule
\multirow{7}{*}{$\widehat{\psi}_{\text{bal}}$} & \texttt{quadprog}   & 0.07 & 0.04 & 0.03 & F    & F     & 0.41 & 0.24 & 0.18 & F    & F    \\
                           & \texttt{qpoases}    & 0.07 & 0.05 &  F & F    & F      & 0.41 & 0.28 & F & F    & F    \\
                           & \texttt{gurobi}     & 0.05 & 0.03 & 0.02 & 0.01 & 0.01   & 0.38 & 0.21 & 0.19 & 0.09 & 0.03 \\
                           & \texttt{Rmosek}      & 0.05 & 0.03 & 0.02 & 0.01 & 0.01    & 0.38 & 0.21 & 0.19 & 0.09 & 0.03 \\
                           & \texttt{osqp}       & 0.05 & 0.03 & 0.02 & 0.01 & 0.01   & 0.39 & 0.21 & 0.19 & 0.10 & 0.03 \\
                           & \texttt{pogs}       & 0.7 & 0.04 & 0.03 & {M}    & {M}     & 0.40 & 0.25 & 0.25 & {M}    & {M}    \\
                           & \texttt{scs}        & 0.11  & 0.10  & 0.09  & 0.05  & 0.02  & 0.39 & 0.25 & 0.20 & 0.13 & 0.10 \\ \bottomrule
\end{tabular}
\caption{RMSE under weak overlap}  \label{tbl:sim-rmse-NL-so}
\end{table}

\clearpage
\newpage

\section*{Appendix D. Time reduction via factorization caching and warm starting} \label{appsec:multi-iter}

\begin{figure}[h!]
\centering
\includegraphics[width=0.71\textwidth]{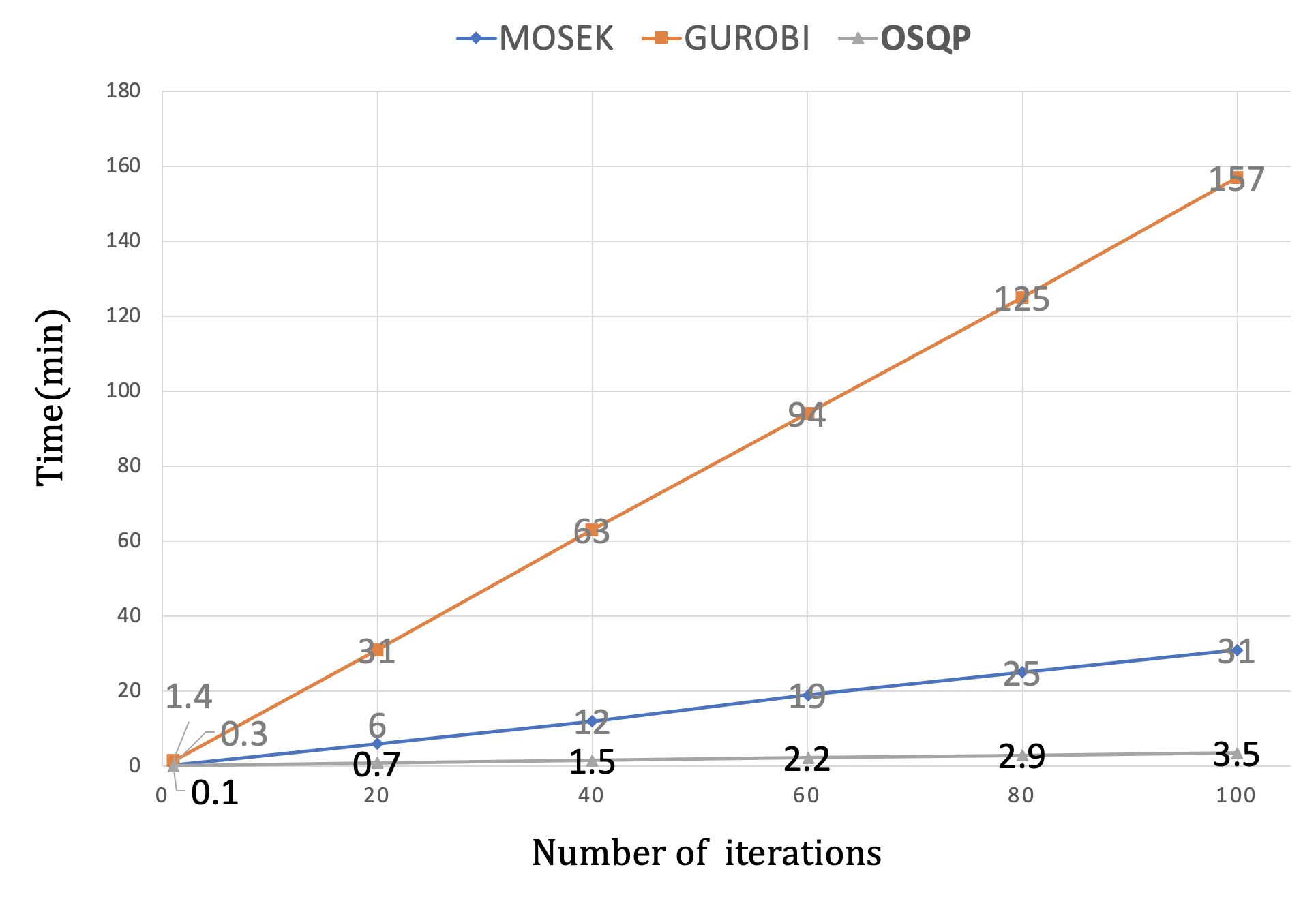}
\caption{Average cumulative CPU time over multiple iterations of SBW calls}
\label{fig:multi-iter}
\end{figure}

Since the vectors $l, u$ are the only parameters that vary in $\delta$ in Algorithm \ref{algo:osqp}, we can speed up repeated OSQP calls via factorization caching and warm starting. Specifically, although the tuning procedure requires a number of iterations, because the coefficient matrix in the relevant linear system is fixed across iterations and does not depend on $\delta$, factorization caching can be used, storing the factorization from the first iteration for use in subsequent ones. Also, because small changes in $\delta$ do not substantially affect weighting solutions, warm starting can be used for added efficiency. To illustrate this computational gain, we solve a sequence of OSQP-based SBW problems using $100$ values of $\delta$ equally spaced between $0.001$ and $0.1$ on the same data, and compare its computing time against two popular commercial solvers \texttt{gurobi} and \texttt{Rmosek} (with the default setting of interior-point methods). Data are generated by the same simulation setup used in Section \ref{sec:simulation} under weak overlap. We measure the cumulative time for each solver at $m=1, 20, 40, 60, 80, 100$ problem instances across different values of $\delta$. We repeat the simulation $100$ times and present their average values in Web Figure \ref{fig:multi-iter}.

For \texttt{gurobi} and \texttt{Rmosek}, the cumulative time increases linearly with the number of iterations as would be expected; i.e., $m$ iterations took roughly $m$ times as long as a single execution. However, $m$ iterations of the OSQP algorithm took much less time than $m \times \text{(the execution time for a single run)}$; in fact, it is on average $2.4$ times faster. This $2.4$-fold time improvement demonstrates the benefits of factorization caching and warm starting via OSQP when implementing SBW.

\clearpage
\newpage

\section*{Appendix E. Complete case study balance table before and after stable kernel balancing}

 \begin{longtable}{p{0.4\linewidth} | p{0.1\linewidth} | p{0.13\linewidth}
 | p{0.1\linewidth} | p{0.1\linewidth} | p{0.1\linewidth}} 
   \hline
  Covariate & For-Profit & Unweighted Controls & Weighted Controls & TASMD Before & TASMD After \\ 
   \hline
 Age at Admission & 79.043 & 79.322 & 79.035 & 0.034 & 0.001 \\ 
   Age 66-74 & 0.369 & 0.360 & 0.367 & 0.019 & 0.005 \\ 
   Age 75-84 & 0.371 & 0.365 & 0.371 & 0.012 & 0.002 \\ 
   Age 85+ & 0.260 & 0.275 & 0.262 & 0.034 & 0.004 \\ 
   Admission Year & 2014.771 & 2014.668 & 2014.778 & 0.039 & 0.002 \\ 
   Admission 2010 & 0.029 & 0.030 & 0.029 & 0.008 & 0.000 \\ 
   Admission 2011 & 0.110 & 0.118 & 0.109 & 0.027 & 0.002 \\ 
   Admission 2012 & 0.112 & 0.117 & 0.112 & 0.016 & 0.001 \\ 
   Admission 2013 & 0.107 & 0.112 & 0.108 & 0.016 & 0.001 \\ 
   Admission 2014 & 0.109 & 0.108 & 0.108 & 0.004 & 0.004 \\ 
   Admission 2015 & 0.111 & 0.108 & 0.111 & 0.010 & 0.000 \\ 
   Admission 2016 & 0.113 & 0.110 & 0.113 & 0.009 & 0.001 \\ 
   Admission 2017 & 0.114 & 0.108 & 0.116 & 0.019 & 0.003 \\ 
   Admission 2018 & 0.099 & 0.098 & 0.100 & 0.003 & 0.003 \\ 
   Admission 2019 & 0.094 & 0.089 & 0.094 & 0.018 & 0.002 \\ 
   Sex (female) & 0.478 & 0.477 & 0.477 & 0.003 & 0.002 \\ 
   Non-Hispanic White & 0.790 & 0.842 & 0.794 & 0.128 & 0.008 \\ 
   Black & 0.083 & 0.078 & 0.082 & 0.019 & 0.004 \\ 
   Hispanic & 0.086 & 0.044 & 0.085 & 0.148 & 0.003 \\ 
   Asian/Pacific Islander & 0.025 & 0.019 & 0.025 & 0.039 & 0.001 \\ 
   Native American & 0.005 & 0.005 & 0.004 & 0.007 & 0.019 \\ 
   Other Race & 0.011 & 0.012 & 0.011 & 0.013 & 0.001 \\ 
   Emergent Admission & 0.837 & 0.838 & 0.836 & 0.002 & 0.003 \\ 
   Anterior ST-elevation MI 1 & 0.069 & 0.077 & 0.069 & 0.032 & 0.002 \\ 
   Other ST-elevation MI & 0.149 & 0.156 & 0.150 & 0.018 & 0.002 \\ 
   History of PCI & 0.150 & 0.159 & 0.152 & 0.024 & 0.004 \\ 
   History of CABG & 0.119 & 0.119 & 0.119 & 0.001 & 0.001 \\ 
   Hx Sepsis & 0.015 & 0.014 & 0.015 & 0.005 & 0.001 \\ 
   Hx Infection & 0.002 & 0.002 & 0.002 & 0.001 & 0.001 \\ 
   Hx Metastatic Cancer & 0.011 & 0.013 & 0.011 & 0.024 & 0.001 \\ 
   Hx Metastatic or Severe Cancer & 0.022 & 0.025 & 0.022 & 0.024 & 0.001 \\ 
   Hx Lung or Severe Cancer & 0.014 & 0.016 & 0.014 & 0.015 & 0.001 \\ 
   Hx Non-metastatic Cancer & 0.029 & 0.034 & 0.030 & 0.027 & 0.003 \\ 
   Hx Cancer, Other & 0.003 & 0.004 & 0.002 & 0.009 & 0.015 \\ 
   Hx Diabetes & 0.376 & 0.371 & 0.375 & 0.010 & 0.002 \\ 
   Hx Diabetes w/ Complications & 0.129 & 0.134 & 0.130 & 0.015 & 0.002 \\ 
   Hx Diabetes w/o Complications & 0.083 & 0.072 & 0.083 & 0.040 & 0.002 \\ 
   Hx Malnutrition & 0.042 & 0.036 & 0.041 & 0.031 & 0.005 \\ 
   Hx Morbid Obesity & 0.677 & 0.695 & 0.679 & 0.037 & 0.003 \\ 
   Hx Liver Disease & 0.016 & 0.017 & 0.016 & 0.006 & 0.000 \\ 
   Hx Dementia & 0.139 & 0.131 & 0.138 & 0.022 & 0.002 \\ 
   Hx Substance Use & 0.117 & 0.113 & 0.117 & 0.012 & 0.000 \\ 
   Hx Major Psychiatric Disorder & 0.018 & 0.017 & 0.019 & 0.010 & 0.001 \\ 
   Hx Cardio-respiratory Failure or Shock & 0.160 & 0.159 & 0.159 & 0.002 & 0.003 \\ 
   Hx Heart Failure & 0.435 & 0.438 & 0.436 & 0.006 & 0.002 \\ 
   Hx AMI & 0.006 & 0.007 & 0.006 & 0.007 & 0.001 \\ 
   Hx Unstable Angina & 0.065 & 0.062 & 0.065 & 0.011 & 0.000 \\ 

   Hx Valvular Disease & 0.156 & 0.191 & 0.159 & 0.095 & 0.006 \\ 
   Hx Hypertensive Heart Disease & 0.021 & 0.016 & 0.021 & 0.035 & 0.000 \\ 
   Hx Hypertension & 0.476 & 0.477 & 0.476 & 0.002 & 0.000 \\ 
   Hx Arrhythmia, Specified & 0.285 & 0.289 & 0.286 & 0.008 & 0.003 \\ 
   Hx Arrhythmia, Other & 0.129 & 0.152 & 0.130 & 0.070 & 0.005 \\ 
   Hx Stroke & 0.007 & 0.007 & 0.005 & 0.003 & 0.022 \\ 
   Hx Cerebrovascular Disease & 0.040 & 0.045 & 0.039 & 0.026 & 0.002 \\ 
   Hx Precerebral Occlusion & 0.032 & 0.036 & 0.032 & 0.025 & 0.000 \\ 
   Hx Peripheral Vascular Disease & 0.058 & 0.062 & 0.059 & 0.014 & 0.001 \\ 
   Hx Lung Fibrosis & 0.017 & 0.019 & 0.017 & 0.019 & 0.001 \\ 
   Hx Asthma & 0.028 & 0.033 & 0.029 & 0.027 & 0.002 \\ 
   Hx COPD & 0.215 & 0.202 & 0.213 & 0.032 & 0.004 \\ 
   Hx Pneumonia & 0.100 & 0.088 & 0.099 & 0.039 & 0.002 \\ 
   Hx Pneumothorax & 0.019 & 0.018 & 0.019 & 0.007 & 0.001 \\ 
   Hx Other Respiratory Disease & 0.064 & 0.070 & 0.065 & 0.023 & 0.003 \\ 
   Hx Eye, Other & 0.010 & 0.012 & 0.011 & 0.015 & 0.003 \\ 
   Hx Renal Failure & 0.364 & 0.359 & 0.364 & 0.009 & 0.000 \\ 
   Hx UTI & 0.089 & 0.078 & 0.088 & 0.039 & 0.004 \\ 
   Hx Male Genital Disorder & 0.087 & 0.092 & 0.088 & 0.018 & 0.004 \\ 
   Hx Skin Ulcer & 0.022 & 0.024 & 0.022 & 0.016 & 0.002 \\ 
   Hx Skin Disease, Other & 0.010 & 0.013 & 0.010 & 0.027 & 0.000 \\ 
   Hx Iatrogenic Complication & 0.035 & 0.037 & 0.035 & 0.010 & 0.001 \\ 
   Hx Trauma & 0.023 & 0.024 & 0.024 & 0.006 & 0.003 \\ 
   \emph{County Hospital Characteristics}  & & & & & \\ 
   Urban Hospital & 0.881 & 0.877 & 0.883 & 0.010 & 0.006 \\ 
   Large Size (501+ Beds) & 0.052 & 0.077 & 0.052 & 0.338 & 0.007 \\ 
   Medium Size (251-100 Beds) & 0.153 & 0.194 & 0.154 & 0.305 & 0.006 \\ 
   Small Size (Under 251 Beds) & 0.795 & 0.729 & 0.793 & 0.442 & 0.009 \\ 
   Resident-to-Bed Ratio & 0.021 & 0.034 & 0.021 & 0.336 & 0.002 \\ 
   Major Teaching (RTB $\geq${0.25}) & 0.026 & 0.039 & 0.026 & 0.195 & 0.006 \\ 
   Minor Teaching (0 $\leq${RTB} $\leq${0.25}) & 0.098 & 0.146 & 0.098 & 0.390 & 0.001 \\ 
   Nonteaching (RTB=0) & 0.876 & 0.815 & 0.876 & 0.407 & 0.001 \\ 
   Nurse-to-Bed Ratio & 0.971 & 1.063 & 0.977 & 0.120 & 0.008 \\ 
   High NTB (NTB$\geq${1.0}) & 0.376 & 0.421 & 0.377 & 0.214 & 0.002 \\ 
   Moderate NTB (0.9 $\leq${NTB} \lt 1.0) & 0.053 & 0.052 & 0.053 & 0.009 & 0.005 \\ 
   Low NTB (NTB$<$0.9) & 0.571 & 0.527 & 0.571 & 0.208 & 0.000 \\ 
   Cardiac Technology Index & 0.921 & 1.022 & 0.921 & 0.263 & 0.000 \\ 
   Cardiac Surgery Provided & 0.246 & 0.280 & 0.246 & 0.234 & 0.004 \\ 
   Coronary Care Unit & 0.411 & 0.480 & 0.412 & 0.330 & 0.005 \\ 
   Cardiac Catheterization Lab & 0.265 & 0.262 & 0.263 & 0.012 & 0.009 \\ 
   All Three Cardiac Services Available & 0.140 & 0.154 & 0.137 & 0.112 & 0.022 \\ 
    \hline
 \end{longtable}